\UseRawInputEncoding

\documentclass[trackchanges,twocolumn,twocolappendix,longbib]{aastex701}
\hypersetup{linkcolor=red,citecolor=green,filecolor=cyan,urlcolor=magenta}
\usepackage{amsmath}

\newcommand{\parder}[2]{ \frac{\partial #1}{\partial #2} }

\def \fexix {Fe\,{\sc xix}}
\def \fexv {Fe\,{\sc xv}}
\def \feix {Fe\,{\sc ix}}

\begin{document}

\title{3D MHD simulations of coronal loops heated via magnetic braiding \\ I. Continuous driving}

\author[orcid=0009-0005-4545-1010,sname='Gabriele Cozzo']{G. Cozzo}
\affiliation{Harvard-Smithsonian Center for Astrophysics, 60 Garden St., Cambridge, MA 02193, USA}
\affiliation{INAF-Osservatorio Astronomico di Palermo, Piazza del Parlamento 1, I-90134 Palermo, Italy}
\email[show]{gabriele.cozzo@cfa@harvard.edu}  

\author[orcid=0000-0002-0405-0668, sname='Paola Testa']{Paola Testa} 
\affiliation{Harvard-Smithsonian Center for Astrophysics, 60 Garden St., Cambridge, MA 02193, USA}
\email{ptesta@cfa.harvard.edu}

\author[orcid=0000-0002-0333-5717, sname='Juan Martinez-Sykora']{J. Martinez-Sykora}
\affiliation{Lockheed Martin Solar and Astrophysics Laboratory, 3251 Hanover St, Palo Alto, CA 94304, USA}
\affiliation{Rosseland Centre for Solar Physics, University of Oslo, P.O. Box 1029 Blindern, N-0315 Oslo, Norway}
\affiliation{Institute of Theoretical Astrophysics, University of Oslo, P.O. Box 1029 Blindern, N-0315 Oslo, Norway}
\affiliation{SETI Institute, 339 Bernardo Ave, Suite 200, Mountain View, CA, 94043, United States}
\email{jmsykora@seti.org} 

\author[orcid=0000-0002-1820-4824, sname='Fabio Reale']{F. Reale}
\affiliation{Dipartimento di Fisica \& Chimica, Universit\`a di Palermo, Via Archirafi 36, 90123 Palermo, Italy}
\affiliation{INAF-Osservatorio Astronomico di Palermo, Piazza del Parlamento 1, I-90134 Palermo, Italy}
\email{fabio.reale@unipa.it} 

\author[orcid=0000-0001-5274-515X, sname='Paolo Pagano']{P. Pagano}
\affiliation{Dipartimento di Fisica \& Chimica, Universit\`a di Palermo, Via Archirafi 36, 90123 Palermo, Italy}
\affiliation{INAF-Osservatorio Astronomico di Palermo, Piazza del Parlamento 1, I-90134 Palermo, Italy}
\email{paolo.pagano@unipa.it} 

\author[orcid=0000-0001-9030-0418, sname='Antonino Franco Rappazzo']{F. Rappazzo}
\affiliation{Dipartimento di Fisica \& Chimica, Universit\`a di Palermo, Via Archirafi 36, 90123 Palermo, Italy}
\affiliation{INAF-Osservatorio Astronomico di Palermo, Piazza del Parlamento 1, I-90134 Palermo, Italy}
\email{antoniofranco.rappazzo@unipa.it} 

\author[orcid=0000-0003-0975-6659, sname='Viggo Hansteen']{V. Hansteen}
\affiliation{Rosseland Centre for Solar Physics, University of Oslo, P.O. Box 1029 Blindern, N-0315 Oslo, Norway}
\affiliation{Institute of Theoretical Astrophysics, University of Oslo, P.O. Box 1029 Blindern, N-0315 Oslo, Norway}
\affiliation{SETI Institute, 339 Bernardo Ave, Suite 200, Mountain View, CA, 94043, United States}
\email{vhansteen@seti.org} 

\author[orcid=0000-0002-8370-952X, sname='Bart De Pontieu']{B. De Pontieu}
\affiliation{Lockheed Martin Solar \& Astrophysics Laboratory, 3251 Hanover St, Palo Alto, CA 94304, USA}
\affiliation{Rosseland Centre for Solar Physics, University of Oslo, P.O. Box 1029 Blindern, N-0315 Oslo, Norway}
\affiliation{Institute of Theoretical Astrophysics, University of Oslo, P.O. Box 1029 Blindern, N-0315 Oslo, Norway}
\email{bdp@lmsal.com}

\author[orcid=0000-0002-9882-1020, sname='Antonino Petralia']{A. Petralia}
\affiliation{INAF-Osservatorio Astronomico di Palermo, Piazza del Parlamento 1, I-90134 Palermo, Italy}
\email{antonino.petralia@inaf.it} 

\author[orcid=0000-0002-9346-1732, sname='Edoardo Alaimo']{E. Alaimo}
\affiliation{Dipartimento di Fisica \& Chimica, Universit\`a di Palermo, Via Archirafi 36, 90123 Palermo, Italy}
\affiliation{INAF-Osservatorio Astronomico di Palermo, Piazza del Parlamento 1, I-90134 Palermo, Italy}
\email{edoardo.alaimo@unipa.it} 

\author[orcid=0009-0000-6975-5127, sname='Federico']{F. Fiorentino}
\affiliation{Dipartimento di Fisica \& Chimica, Universit\`a di Palermo, Via Archirafi 36, 90123 Palermo, Italy}
\affiliation{INAF-Osservatorio Astronomico di Palermo, Piazza del Parlamento 1, I-90134 Palermo, Italy} 
\email{federico.fiorentino@unipa.it} 

\author[orcid=0000-0001-7834-5760, sname='Fabio D'anca']{F. D'Anca}
\affiliation{INAF-Osservatorio Astronomico di Palermo, Piazza del Parlamento 1, I-90134 Palermo, Italy}
\email{fabio.danca@inaf.it} 

\author[orcid=0000-0002-4725-357X, sname='Luisa Sciortino']{L. Sciortino}
\affiliation{Dipartimento di Fisica \& Chimica, Universit\`a di Palermo, Via Archirafi 36, 90123 Palermo, Italy}
\affiliation{INAF-Osservatorio Astronomico di Palermo, Piazza del Parlamento 1, I-90134 Palermo, Italy}
\email{luisa.sciortino@unipa.it} 

\author[orcid=0000-0002-3843-1637, sname='Michela Todaro']{M. Todaro}
\affiliation{INAF-Osservatorio Astronomico di Palermo, Piazza del Parlamento 1, I-90134 Palermo, Italy}
\email{michela.todaro@inaf.it} 

\author[orcid=0000-0003-0318-3546, sname='Ugo Lo Cicero']{U. Lo Cicero}
\affiliation{INAF-Osservatorio Astronomico di Palermo, Piazza del Parlamento 1, I-90134 Palermo, Italy}
\email{ugo.locicero@inaf.it} 

\author[orcid=0000-0002-3188-7420, sname='Marco Barbera']{M. Barbera}
\affiliation{Dipartimento di Fisica \& Chimica, Universit\`a di Palermo, Via Archirafi 36, 90123 Palermo, Italy}
\affiliation{INAF-Osservatorio Astronomico di Palermo, Piazza del Parlamento 1, I-90134 Palermo, Italy}
\email{marco.barbera@inaf.it} 

\begin{abstract}

The nature and detailed properties of the heating of the million-degree solar corona are important issues that are still largely unresolved. Nanoflare heating might be dominant in active regions and quiet Sun, although direct signatures of such small-scale events are difficult to observe in the highly conducting, faint corona.
The aim of this work is to test the theory of coronal heating by nanoflares in braided magnetic field structures.
We analyze a 3D MHD model of a multistrand flux tube in a stratified solar atmosphere, driven by \textcolor{black}{twisting motions} at the boundaries. 
We show how the magnetic structure is maintained at high temperature and for an indefinite time, by intermittent episodes of local magnetic energy release due to reconnection. We forward-modelled optically thin emission with SDO/AIA and MUSE and compared the synthetic observations with the intrinsic coronal plasma properties, focusing on the response to impulsive coronal heating. Currents build up and their impulsive dissipation into heat are also investigated through different runs.
In this first paper, we describe the proliferation of heating from the dissipation of narrow current sheets in realistic simulations of braided coronal flux tubes at unprecedented high spatial resolutions.

\end{abstract}


\section{Introduction}

The million-degree solar corona owes its extreme temperature to the continuous driving of energy from the coronal magnetic field to its tenuous plasma \citep{Alfven1947,parker1988nanoflares,testa2015,klimchuk2015key,testa2023solar}. In  \textcolor{black}{Extreme Ultraviolet (EUV)} and X-ray images, this interaction manifests itself as bundles of bright, arch-shaped strands, the coronal loops, that outline closed magnetic flux tubes linking opposite-polarity regions in the photosphere \citep{vaiana1973identification,reale2014coronal}. Identifying the physical processes that maintain such hot structures remains one of the central challenges of solar physics \citep{grotrian1939sonne,edlen1943deutung,peter2014discovery}.

Coronal loops remain bright for hours, much longer than their characteristic cooling times \citep{rosner1978dynamics}. For most of their lifetime, they can therefore be described as systems in equilibrium \citep{landini1975loop}. The rise, steady, and decay phases of their emission imply that the average heating rate slowly grows, plateaus, and declines.
Observations revealed that coronal loops are also highly dynamic and structured, both spatially and thermally. Fuzziness seen in harder X‑ray bands and hotter spectral lines was interpreted as evidence that each loop is a bundle of thin, unresolved strands heated in pulses \citep{guarrasi2010coronal}.
To explain both the long‑lived brightness and the fine structure, several authors have argued that loops are not in true equilibrium. Instead, they are filamented and continuously cool from a hotter state while undergoing repeated heating episodes \citep{cargill2004nanoflare}.

In the classic picture proposed by \citet{parker1988nanoflares}, random foot-point shuffling braids coronal field lines into ever more tangled configurations. At sufficiently small spatial scales (strands of the order of $10$–$100\,\mathrm{km}$ wide, according to both observations and models; \citealt{beveridge2003model,klimchuk2008highly,vekstein2009probing}) intense current sheets must form. Reconnection across those sheets releases packets of magnetic energy of order $10^{24}\,$erg, the so-called ``nanoflares'' \citep{parker1988nanoflares}, believed to be the small-energy limit of a ``syndrome'' \citep{hudson2010observations} of observed phenomena with energies following a frequency power-law of index $\sim -1.8$. \citep{drake1971characteristics, dennis1985solar, crosby1993frequency}.
Individual nanoflares are often difficult to isolate observationally  \citep[e.g.,][]{testa2013observing,testa2014evidence,testa2020moss,cho2023}, as thermal conduction rapidly spreads the heat along the reconnected field lines \citep[e.g.,][]{gudiksen2005ab}. 

Ab-initio and boundary-driven 3D MHD simulations offer complementary views of how energy is transported into and dissipated within coronal loops. Ab-initio models that extend from the upper convection zone through the photosphere, chromosphere, and transition region self-consistently produce loop turbulence, current-sheet formation, and intermittent heating; e.g., \citet{breu2022solar} model a straightened flux tube with MURaM \citep{vogler2005simulations,rempel2016extension} and recover transient bright strands in synthesized AIA/XRT emission \citep{lemen2012atmospheric,golub2008x}. Boundary-driven studies similarly reveal that complex footpoint motions promote the proliferation and dissipation of thin current sheets and bursty energy release, with heating characteristics that depend on the frequency and amplitude of the driver \citep{howson2022effects}. Taken together, these simulations link together magnetic braiding, turbulence, and reconnection, establishing a consistent mesoscale picture of coronal-loop structuring and heating.

\textcolor{black}{The merging of two or more twisted flux ropes is also considered a potential mechanism for releasing stored magnetic energy, leading to both large-scale flares and, on smaller scales, coronal heating by nanoflares \citep[e.g.,][]{gold1960origin, melrose1997solar, kondrashov1999three}.  Reconnection and energy release during flux rope mergers have been both modelled \citep[e.g.,][]{linton2001reconnection, kliem2014slow} and observed in various eruptive events \citep{liu2020magnetic}.
Recent 2.5D simulations by \cite{sen2025merging} using the MPI-AMRVAC code revealed that the interaction and merging of flux ropes can also give rise to nanojet-like \citep{antolin2021reconnection, 2025A&A...695A..40C} ejections.}

Lare3D (\citealt{arber2001staggered}) simulations show that in a multistrand, non-potential coronal loop, a single kink-unstable thread can trigger an avalanche of reconnection across neighboring strands, even if they are individually sub-critical \citep{tam2015coronal,2016mhhoodd}. Once a strand exceeds the kink twist threshold, a helical current sheet forms, fragments, and dissipates, producing nanoflare-like bursts \citep{hood2009coronal,baty1996electric}. The resulting heating is intermittent, with short spikes rather than steady dissipation, providing a triggered mechanism for loop energization. Energy injected by photospheric motions is released via viscous and Ohmic dissipation during relaxation and, under continuous driving, through repeated impulsive, localized events \citep{reid2018coronal,reid2020coronal}. Whether further avalanche onsets are suppressed by field disorder remains debated \citep{rappazzo2013field}. Following three neighboring threads twisted by a footpoint driver, \citet{reid2018coronal,reid2020coronal} confirmed that the destabilization of a single “node” can trigger the release of magnetic energy across all three and sustain bursty heating; with ongoing twisting, the corona approaches a statistically steady state in which intermittent pulses balance radiative and conductive losses. 

PLUTO \citep{mignone2007pluto} simulations described in \cite{cozzo2023coronal} extended previous MHD avalanche models, accounting for a fully stratified upper-atmosphere (chromosphere, transition region, and corona, modelled through Spitzer thermal conduction, \citealt{spitzer1953transport}, optically thin radiative losses, and gravity), allowing for detailed forward modelling of emission in optically thin EUV lines \citep{cozzo2024coronal}. Recently, \cite{johnston2025self} showed that twisting-induced MHD-avalanches can lead to strong and compact events of energy release (nanoflare storms) that may be responsible for the bright loops that stand out in the corona against the diffuse EUV emission.

The forthcoming Multi-slit Solar Explorer \citep[MUSE, ][]{de2019multi} will address coronal-loop heating through high-cadence, high-resolution EUV spectroscopy, delivering simultaneous constraints on brightness, Doppler shifts, and non-thermal line widths over a broad temperature range \citep[MUSE, ][]{de2022probing}.
This will provide new insights on how magnetic reconnection operates within coronal magnetic-flux strands, the fundamental elements of loops and active regions, and will open new avenues for diagnosing coronal heating. In preparation, a more complete theoretical and numerical framework for DC heating via component 3D reconnection \citep{hesse1988theoretical, schindler1988general} in  braided fields is required and must be tested in a realistic coronal environment that includes chromospheric and photospheric coupling, layers that supply both Poynting flux and mass to the corona through nonlinear MHD processes \citep{parnell2012contemporary}. 

For instance, building on \citet{cozzo2023coronal} setup, \citet{cozzo2024coronal}
sampled at the MUSE expected pixel scale, cadence, and temperature response.
They found that MUSE could resolve fine structures from tube fragmentation, including footpoints evaporation upflows (Fe IX \textcolor{black}{at 171 $\AA$}), bulk loop brightening with alternating filamentary Doppler patterns (Fe XV \textcolor{black}{at 284 $\AA$}), and transient brightenings of folded sheet-like structures bounding unstable strands (Fe XIX \textcolor{black}{at 108 $\AA$}). In a following work, \citet{2025A&A...695A..40C} developed diagnostics for reconnection outflows in hot coronal loops, widely interpreted as signatures of ubiquitous small-angle component reconnection in the solar corona \citep{antolin2021reconnection}. They show that the MUSE spectrograph can isolate EUV lines selectively sensitive to hot plasma, offering clear advantages over previous instruments due to its high-temperature sensitivity and its ability to detect the predicted bidirectional Doppler shifts. 
Overall, the predicted signatures indicate that MUSE can pinpoint key dynamics of ignition and reconnection, providing discriminating diagnostics of the underlying heating processes.

The next step is to address the long temporal brightening of observed coronal loops and the role of DC heating in coronal close magnetic flux tubes. With this aim, in this paper, we investigated the post-avalanche phase of the coronal flux tubes in a stratified solar atmosphere. Specifically, we pursued mesoscale simulations that track a few strands, trading domain extent for the resolution needed to capture current-sheet formation, thinning, and reconnection outflows, physics that is generally addressed only in simpler reduced-MHD or 2.5D studies \citep[e.g., ][]{rappazzo2013field}, while incorporating key coronal ingredients (e.g., anisotropic conduction and optically thin radiative losses) generally used in large scale AR simulation, to enable observation-driven validation. A continuous footpoints driver \citep{reid2018coronal} powers a nearly constant Poynting flux into the corona. We consider a time scale sufficiently long as to reach a steady state. We address the energy build-up of the already fragmented magnetic field and its intermittent, impulsive dissipation within narrow current sheets.

In section \ref{sec:model} we describe the model, in section \ref{sec:results} we report on the evolution of the coronal loop on a global scale, on the reconnection processes occurring in the post-avalanche phase of the loop evolution, and eventually on forward modeling results of EUV emission in AIA channels and MUSE spectral lines \citep{de2022probing} and on the role of magnetic diffusivity through the analysis of different runs with different numerical parameters. We draw our conclusions in section \ref{sec:conclusions}.

\section{Model}
\label{sec:model}

\begin{figure*}[tp]
   \centering
\includegraphics[width=0.9\hsize]{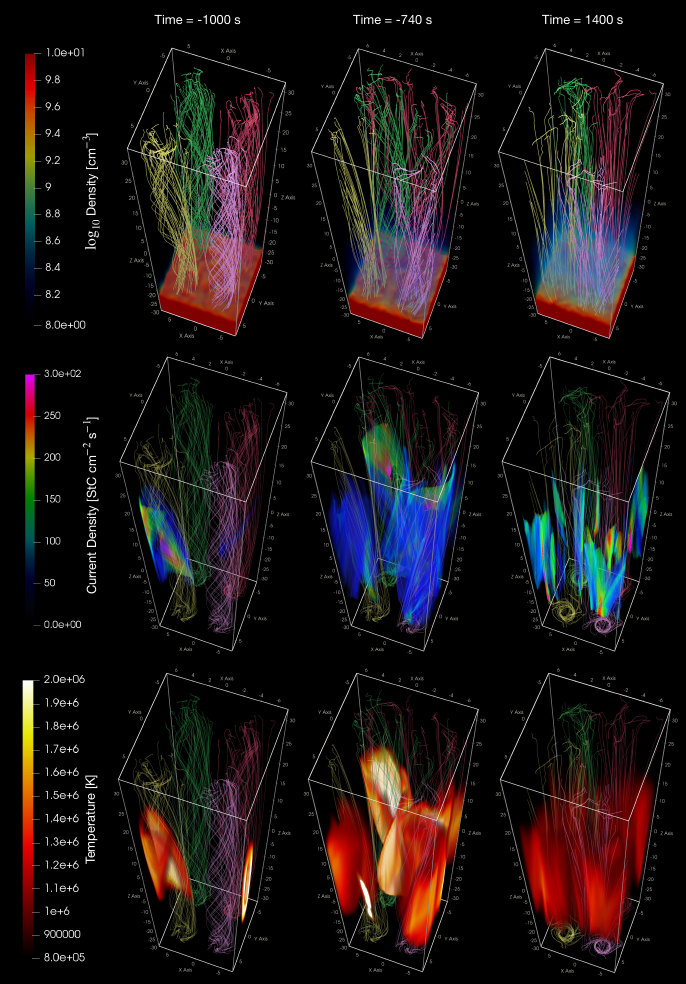}
  \caption{3D rendering of a simulation of a coronal loop in a box at three time snapshots times (before the avalanche: $t = -1000\,\mathrm{s}$; during the avalanche: $t=-740\,\mathrm{s}$; after the avalanche: $t = 1400\,\mathrm{s}$). Coloured field lines show the four magnetic field bundles subjected to continuous footpoints twisting. The first row shows the density stratification from the lower footpoint. The second row shows the current density magnitude in the corona. The third row shows the temperature distribution.} 
  \label{Fig:paper_1_3D_view}
\end{figure*}

We used a model of a magnetized solar atmosphere in a 3D cartesian box where interacting and twisted coronal loop strands are braided at sub-arcsec scales, similar to the work presented by \cite{cozzo2023asymmetric} and \cite{reale20163d}.

The simulations are performed using the PLUTO code \citep{mignone2007pluto}, a modular Godunov-type code for astrophysical plasmas. The evolution of the plasma and magnetic field in the box is described by solving the full, time-dependent MHD equations for a \textcolor{black}{resistive}, fully ionized plasma, including gravity (for a curved loop), thermal conduction (including the effects of heat flux saturation), radiative losses from
an optically thin plasma, and an anomalous magnetic diffusivity:
\begin{align}
&\parder{\rho}{t} + \mathbf{\nabla} \cdot (\rho \vec v) = 0, \label{Eq:PLUTO_MHD_1}\\
&\parder{\rho \vec v}{t} + \mathbf{\nabla} \cdot (\rho \vec v \vec v) = - \nabla \cdot (\mathbf{I} P + \mathbf{I} \frac{B^2}{8 \pi} - \frac{\vec B \vec B}{8 \pi}) + \rho \vec g, \label{Eq:PLUTO_MHD_2}\\
&\parder{B}{t} - \nabla \times (\vec v \times \vec B) = - \eta \nabla^2 \vec B, \label{Eq:PLUTO_MHD_3}\\
&\parder{}{t} \left( \frac{B^2}{8 \pi} + \frac{1}{2} \rho v^2 + \rho \epsilon + \rho g h\right) + \nabla \cdot [ \frac{c}{4 \pi} \vec E \times \vec B + \frac{1}{2}\rho v^2 \vec v \\ & + \frac{\gamma}{\gamma-1} P \vec v + \vec F_{\mathrm{c}} + \rho g h \vec v ] = Q(T), \nonumber \label{Eq:PLUTO_MHD_4} \\
& P = (\gamma -1) \rho \epsilon = \frac{2 k_B}{\mu m_H} \rho T, 
\end{align}
where $t$ is the time; 
$\rho$ is the mass density; 
$\vec{v}$ is the plasma velocity; $P$ is the thermal pressure; 
$\vec{B}$ is the magnetic field; $\vec E$ is the electric field;
$\vec{g}$ 
is the gravity acceleration vector for a curved loop;  $\mathbf{I}$ is the
identity tensor;
$\epsilon$ is the internal energy; 
$\eta$ is the magnetic
diffusivity;  
$T$ is the temperature; 
$\vec{F}_{\mathrm{c}}$ is the thermal conductive flux; $Q(T)$ includes the radiated energy losses and heating; $m_H$ is the hydrogen mass density; $k_B$  is the Boltzmann constant;
$\mu = 1.265$ is the mean ionic weight \citep[relative to a proton and assuming coronal elemental abundances;][]{anders1989abundances}; 

The differential equations are numerically integrated with the MHD module available in PLUTO \citep{mignone2007pluto}, configured to compute intercell fluxes with the Harten-Lax-Van Leer approximate Riemann solver \citep{roe1986characteristic}, while second-order accuracy in time is achieved using a Runge-Kutta scheme. A Van Leer limiter \citep{sweby1985high} for the primitive variables is used. 
The solenoidal condition is maintained with the constrained transport approach \citep{balsara1999staggered}.

\textcolor{black}{The equations are solved over a stratified, magnetized solar atmosphere, featuring a high-beta chromosphere, a tenuous coronal environment, and a narrow transition region, and threaded by multiple magnetic loop strands. Each strand is modeled as a straight coronal flux tube connecting two chromospheric layers, positioned at opposite ends of the computational box. The tube is much longer than it is wide, and except for the loop-aligned gravity, curvature effects are neglected. The chromospheric layers act as distant, independent loop footpoints.}

We consider four interacting magnetic flux tubes, each with length $62\,\mathrm{Mm}$, aspect ratio $\sim 10$, and initial coronal temperature $\sim 1\mathrm{MK}$. \textcolor{black}{The coronal part of the strands is $50\,\mathrm{Mm}$ long.} The computational box is a 3D Cartesian grid, $-x_M < x < x_M$, $-y_M < y < y_M$, and $-z_M < z < z_M$, where $x_M = y_M = 7\,\mathrm{Mm}$, and $z_M = 31\,\mathrm{Mm}$, with a staggered grid, uniform along $\hat x$ and $\hat y$, with $\Delta x = \Delta y \sim 30\,\mathrm{km}$.
Along $\hat z$ we use a non-uniform grid, with high resolution ($\Delta z \sim 25 \,\mathrm{km}$) in the chromosphere and TR, while $\Delta z$ logarithmically increases with height in the corona \textcolor{black}{up to $100\,\mathrm{km}$}.   
The flux tubes extend along the $z$ direction and connect chromospheric layers placed at the top and bottom in this configuration.

To approach a semicircular configuration typical of coronal loops, 
we consider the gravity of the curved flux tube as follows \citep{reale20163d, cozzo2023coronal}:
\begin{equation}
g(z) \hat{\mathbf{z}} =
\begin{cases}
 g_{\odot} \sin{\left(\pi \frac{z}{L}\right)} \,\hat{\mathbf{z}}, & |z| \le 25\,\mathrm{Mm}, \\
 g_{\odot} \, \mathrm{Sign}\,(z) \,\hat{\mathbf{z}} & |z| > 25\,\mathrm{Mm},
\end{cases}
\end{equation}
where $ g_{\odot}  = \frac{G M_{\odot}}{R_{\odot}^2}$ is the gravitational acceleration at the solar surface.

The thermal conductive flux is defined as follows:
\begin{align}
& \vec{F}_c = \frac{F_{\mathrm{sat.}}}{F_{\mathrm{sat.}} + |F_{\mathrm{class.}}|} \vec{F}_{\mathrm{class.}}, \\
& \vec{F}_{\mathrm{class.}} = - k_{\parallel} \hat{\vec{b}} ( \hat{\vec{b}} \cdot \mathbf{\nabla} T) - k_{\perp} \left[ \mathbf{\nabla} T - \hat{\vec{b}} ( \hat{\vec{b}} \cdot \mathbf{\nabla} T) \right], \\
& F_{\mathrm{sat.}} = 5 \phi \rho c^3_{\mathrm{iso.}},
\end{align}
where, $k_{\parallel} = 9.2 \times 10^{-7} T^{\frac{5}{2}}$ and
$k_{\perp} = K_{\perp} = 5.4 \times 10^{-16} \rho^2 / (B^2T^{\frac{1}{2}})$, $c_{\mathrm{iso.}}$ is the isothermal sound speed; $\phi = 0.9$ is a dimensionless free parameter; $\hat{\vec{b}} = \vec B / B$ is the magnetic field unit vector; and $F_{\mathrm{sat.}}$ is the saturated flux.

We account for optically thin radiative cooling in the corona with a rate per unit volume:
\begin{equation}
    Q(T) = - \Lambda(T) n_e n_H + H_0,
\end{equation}
where $\Lambda (T)$ are the radiative rates per unit emission measure, while $n_H$ and $n_e$  are the hydrogen and electron number densities, respectively \textcolor{black}{(see appendix \ref{sec:appendix})}. The volumetric heating rate $H_0 = 4.3 \times 10^{-5}\,\mathrm{erg}\,\mathrm{cm}^{-3}\,\mathrm{s}^{-1}$ is only used in the pre-avalanche phase to balance the initial energy losses and keep the loop in thermal equilibrium, and is switched off when the avalanche starts as the heating released by then is much higher.
The optically thin radiative losses $\Lambda (T)$ values from the CHIANTI v.~7.0 database \citep{landi2013prominence}, assuming coronal element abundances 
\citep{feldman1992elemental}. To describe properly the plasma flows across the transition region \citep{bradshaw2013influence}, 
we modelled the TR with an Adaptive Conduction method \citep[TRAC;][]{johnston2019fast,johnston2020modelling,johnston2021fast}.

We considered an anomalous plasma resistivity \citep{hood2009coronal,reale20163d, cozzo2023coronal} that is switched on only in the corona and TR (i.e. above $T_{\mathrm{cr}} = 10^4\,\mathrm{K}$) where the magnitude of the current density exceeds a threshold:
\begin{equation}
\eta =
    \begin{cases}
    \eta_0 & |J| \ge J_{\mathrm{cr}}  \text{ and } T \ge T_{\mathrm{cr}} \\
    0  & |J| < J_{\mathrm{cr}} \text{ or } T < T_{\mathrm{cr}}
    \end{cases},
    \label{Eq:anomalus_diffusivity}
\end{equation}
where $\eta_0 = 10^{11}\,\mathrm{cm}^{2}\,\mathrm{s}^{-1}$  and $J_{\mathrm{cr}} = 300\,\mathrm{StC}\,\mathrm{cm}^{-2}\,\mathrm{s}^{-1}$ (allowing the build-up of strong currents in narrow current sheets and, consequently, their impulsive dissipation into heating). The minimum heating rate above the threshold is $H = \eta_0 (4 \pi |J_{\mathrm{cr}}|/c)^2 \approx 8 \times 10^{-5}\,\mathrm{erg}\,\mathrm{cm}^{-3}\,\mathrm{s}^{-1}$.  

Each flux tube is rooted in the chromosphere, treated as a simple, thermally conducting, mass reservoir with no radiative losses. The magnetic field has a maximum strength of a few hundred Gauss in the chromosphere, and it expands into the corona to maintain total pressure equilibrium. The magnetic field has an amplitude of about 20 G in the corona and is line-tied to the upper and lower boundaries of the domain. 
\textcolor{black}{In the model, magnetic tapering within the chromosphere results from the relaxation of an initially axial field, stronger around each flux strand \citep{guarrasi2014mhd}. The new equilibrium state is then used as initial condition of the simulation. The initial atmosphere has peak coronal temperature of $8 \times 10^5\,\mathrm{K}$, rapidly decreasing through the transition region to chromospheric temperature ($\sim 10^4\,\mathrm{K}$). Plasma density and pressure distributions are also analogous to those shown in \cite{guarrasi2014mhd}, \cite{reale20163d}, and \cite{cozzo2023asymmetric}.}

\textcolor{black}{The boundary conditions (BCs) are periodic at $x = \pm x_M$ and $y = \pm y_M$. Reflecting BCs for the velocity were set at $z = \pm z_M$. The magnetic field is line-tied to the upper and lower boundaries (representing the dense, highly conducting chromosphere) through equatorially symmetric BCs. Specifically, magnetic footpoints are anchored to these boundary planes, although field lines can slip along them.}

\textcolor{black}{Rotational motions at the upper and lower footpoint boundaries} twist the flux tubes (similar to the setup of \citealt{cozzo2023coronal} and \citealt{reale20163d}). The angular velocity $\omega(r)$ is constant (rigid-body-like) in an inner circle around the central axis of each strand, and then decreases linearly in an outer annulus \citep{reale20163d}: 
\begin{equation}
    v_{\phi} = \omega \, r \, \left[1 + 0.2 \sum_{i=1}^6 \sin{(\phi \alpha_i)} \, \sin{\left(\frac{r}{r_{\mathrm{max}}} \alpha_i \right)} \phi\right]
\end{equation}
where $\alpha_i$ are random numbers between 0 and 30, different for each strand, and:
\begin{equation}
    \omega = \omega_0 \times
    \begin{cases}
    1 & r < r_{\mathrm{max}}   \\
    (2 r_{\mathrm{max}} - r)/r_{\mathrm{max}} & r_{\mathrm{max}} < r < 2 r_{\mathrm{max}}    \\
    0 & r > 2 r_{\mathrm{max}}
    \end{cases}
    \label{eq:angular_vel}
\end{equation}
\textcolor{black}{The four rotating regions have the same radius ($r_{\mathrm{max}} \sim 1\,\mathrm{Mm}$), but in one of them the twisting velocity is 10 \% higher than in the others, reaching a maximum amplitude of $v_{\mathrm{max}} = 2.2\,\mathrm{km/s}$ ($\omega_0 = v_{\mathrm{max}}/r_{\mathrm{max}}$). This value corresponds to the peak rotational speed among the imposed footpoint drivers and is consistent with typical photospheric or chromospheric vortex motions observed in active regions \citep[e.g.,][]{Bonet2008, Galsgaard1996}. Such relatively slow rotational motions are commonly used in numerical models to emulate DC-type magnetic stressing (e.g., \citealt{howson2022effects, johnston2025self}), where the coronal field is gradually braided by persistent photospheric motions. In contrast, AC heating scenarios assume much faster, wave-like perturbations that excite MHD oscillations in the corona.}

\textcolor{black}{We run the simulation for a total duration of 6000 s, from t = -2000 s to t = 4000 s. The earlier evolution of the system, including the onset and development of the MHD avalanche process, has been described in detail in previous studies \citep{cozzo2023coronal, cozzo2024coronal}. In the present work, we focus instead on the subsequent phase driven by the continuous footpoint motions. We define the start of the analyzed time interval (
t = 0 s) immediately after the pronounced peaks in temperature, velocity, and current density associated with the avalanche (see Figure \ref{Fig:paper_1_time_evolution} in the following section). Consequently, negative times in the figures refer to the preceding evolution of the coronal loops, which ultimately leads to the kink instability and the global avalanche event.}

\section{Results}
\label{sec:results}
\subsection{Coronal loop evolution}
\label{sec:loop_evolution}

\begin{figure*}[tp]
   \centering
\includegraphics[width=\hsize]{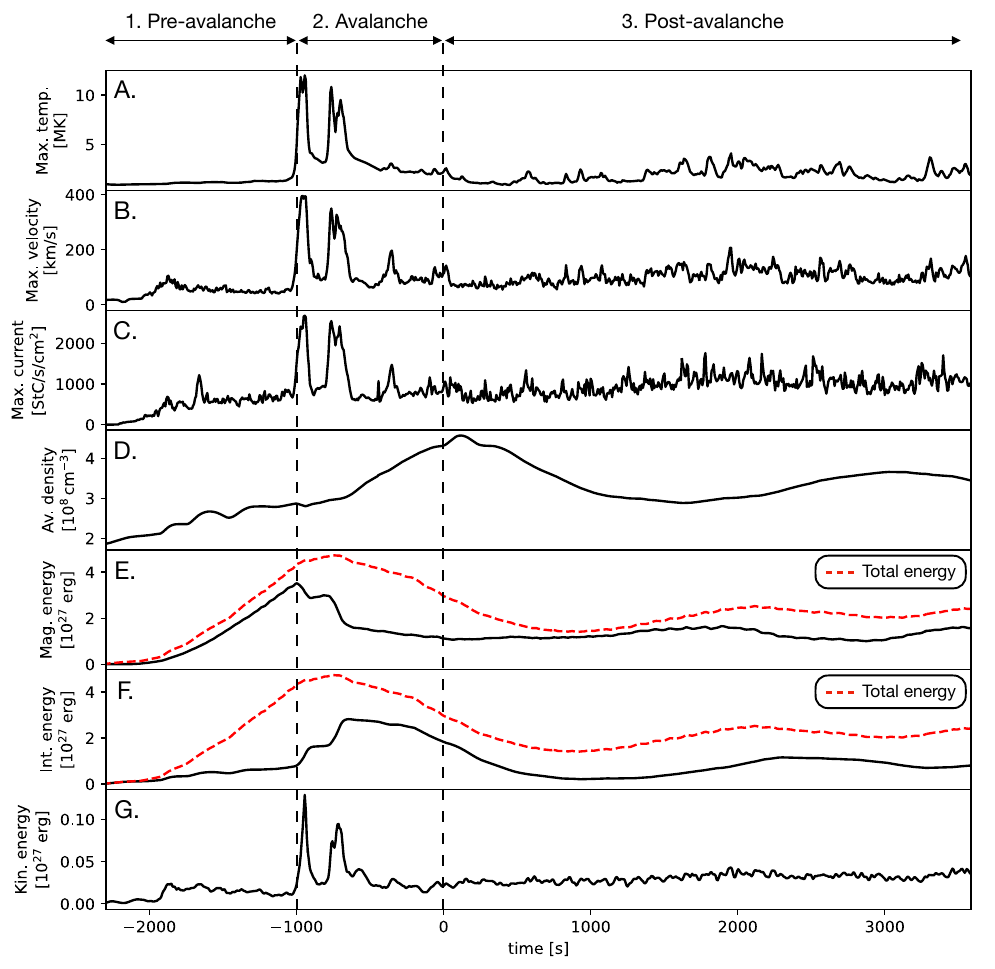}
  \caption{Long-term evolution of the bundle of four flux tubes. From top to bottom: Maximum temperature, maximum velocity, and maximum current density, average coronal plasma density ($T > 10^6\,\mathrm{K}$) , total magnetic, internal, and kinetic energy as functions of time.} 
  \label{Fig:paper_1_time_evolution}
\end{figure*}

\begin{figure*}[h!]
   \centering
\includegraphics[width=\hsize]{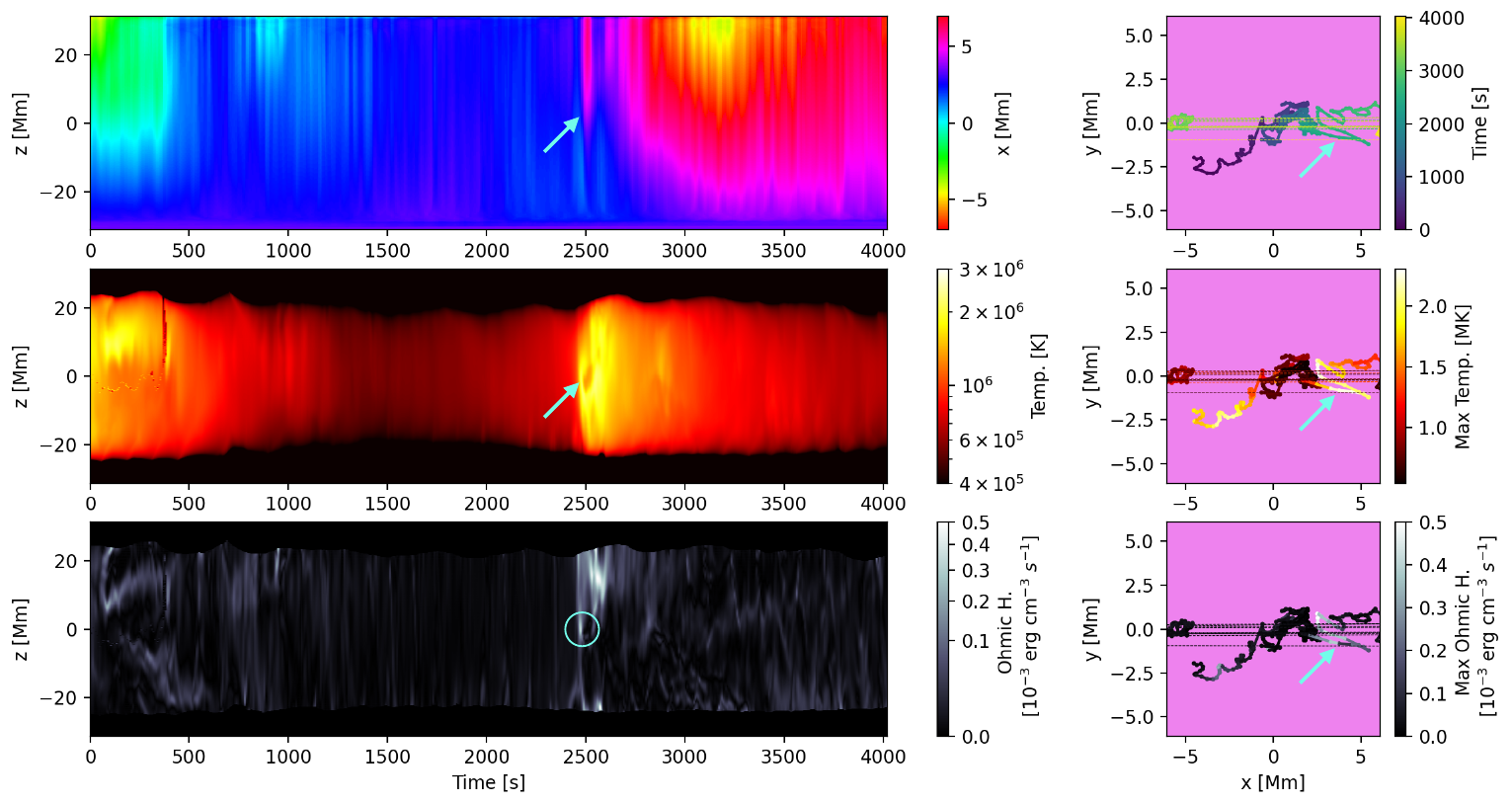}
  \caption{Evolution of a field line and the neighbouring plasma. 
  First row: x-position of the field line along the loop length ($z = 0\,\mathrm{Mm}$ is the apex) as function of time (left), trajectory of the upper footpoint motion (the bottom footpoint moves following the footpoints driver).
  Second row: distribution of the plasma temperature along the field line (left) and its maximum value (right).
  Third row: distribution of the ohmic heating along the field line (left) and its maximum value (right). Cyan arrows and circle highlight a reconnection event.} 
  \label{Fig:paper_1_field_line_evolution}
\end{figure*}

\begin{figure}[h!]
   \centering
\includegraphics[width=\hsize]{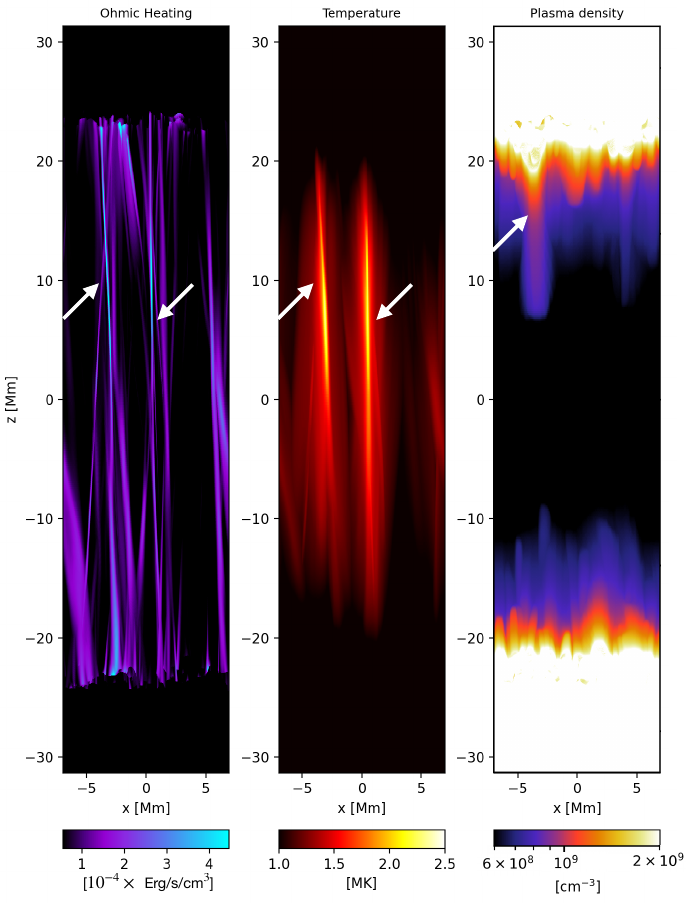}
  \caption{Current sheets along the flux tube. Maps of the distribution of maximum ohmic heating and temperature (along the $\hat y$ direction), and a cut of the plasma density across the box at time $\sim 3000\,\mathrm{s}$ after the onset of the instability. White arrows in the first and second panels point at the location of two current sheets while in the third panel an upflow of plasma evaporating from the TR is shown.} 
  \label{Fig:paper_1_physics_properties}
\end{figure}

\begin{figure*}[h!]
   \centering
\includegraphics[width=\hsize]{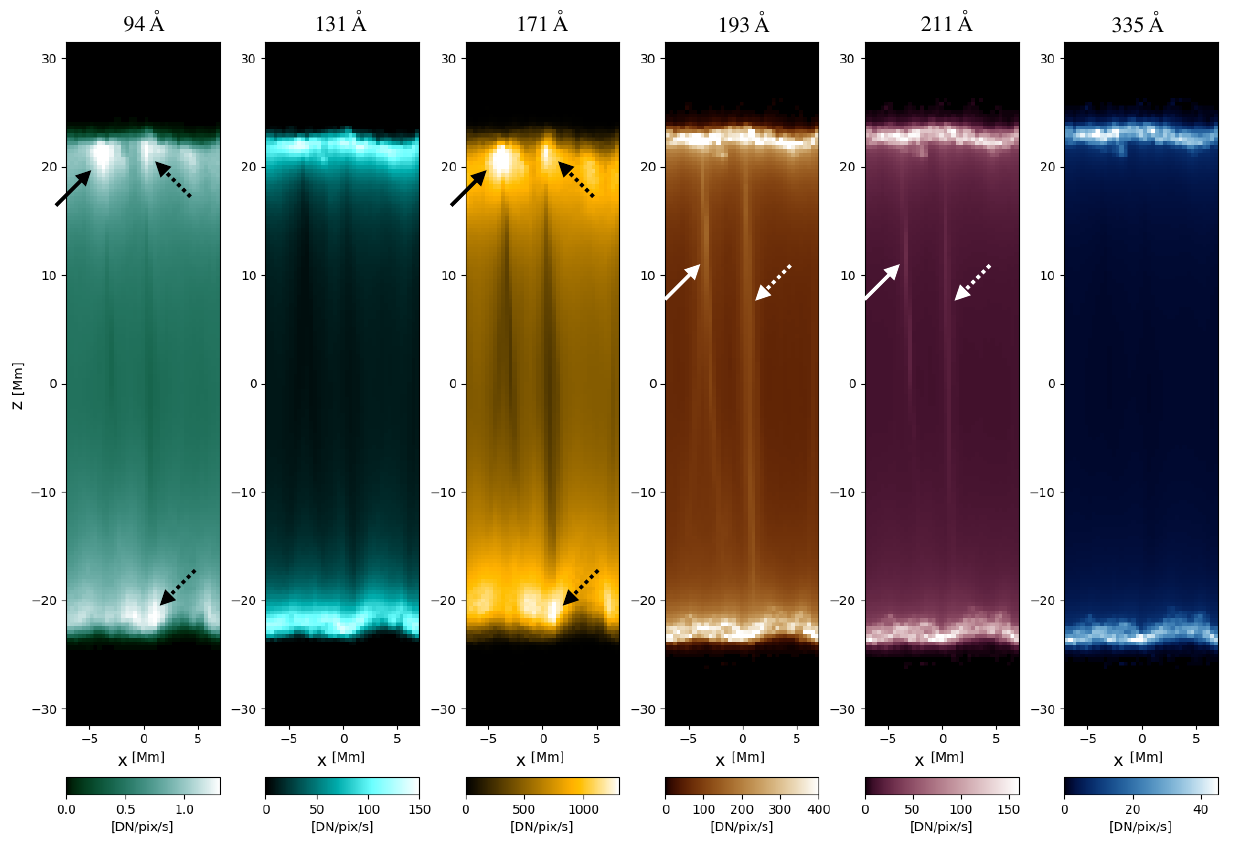}
  \caption{AIA synthetic maps, from the PLUTO 3D MHD simulation of the  multi-threaded magnetic flux tube discussed in Section \ref{sec:model}, at time $\sim 3000\,\mathrm{s}$ after the onset of the instability and with the loop as in off-limb configuration (LoS along the $\hat y$ direction). From the left, intensity in the $94\,\AA$, $131\,\AA$, $171\,\AA$, $193\,\AA$, $211\,\AA$, and $335\,\AA$ channels, respectively. Arrows indicate the brightenings due to impulsive heating events. Movie I shows the temporal evolution of the six panels, across the six optically thin channels of AIA.} 
  \label{Fig:paper_1_AIA_maps}
\end{figure*}

\begin{figure*}[h!]
   \centering
\includegraphics[width=\hsize]{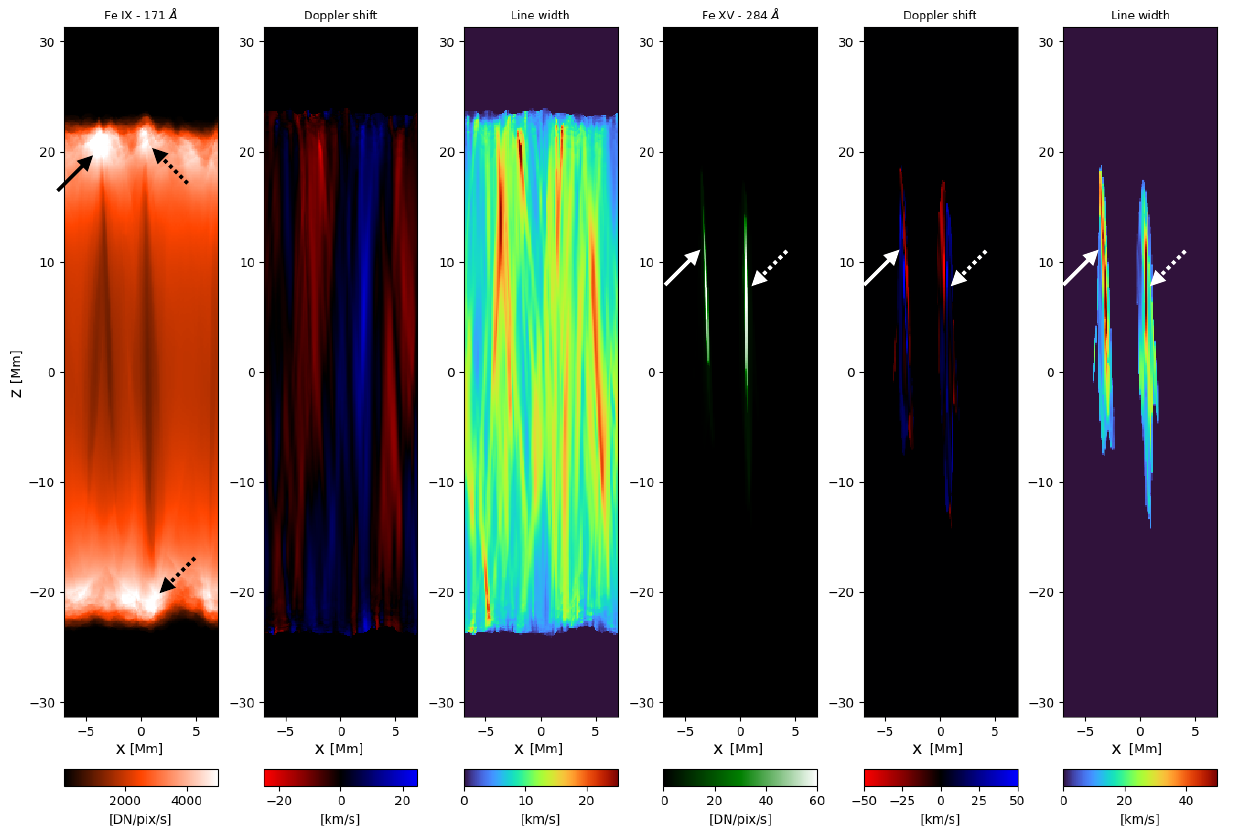}
  \caption{MUSE synthetic maps, from the PLUTO 3D MHD model of the  multi-threaded magnetic flux tube discussed in Section \ref{sec:model}. Here we show the tube structure at time $\sim 3000\,\mathrm{s}$ after the onset of the instability and with the loop as in off-limb configuration (LoS along the $\hat y$ direction). From the left, the panels shows the intensity, Doppler shifts, and non-thermal line broadenings of the \feix\  171 \AA\ and \fexv\ 284 \AA\ emission lines. Arrows as in Fig. \ref{Fig:paper_1_AIA_maps}. Movie II shows the temporal evolution of the six panels, across the Fe IX and Fe XV channels of MUSE.} 
  \label{Fig:paper_1_MUSE_maps}
\end{figure*}

At the beginning (t = -2000 s) the coronal loop is empty ($\rho < 2 \times 10^8\,\mathrm{cm}^{-3}$) and cold ($T < 1\,\mathrm{MK}$). The footpoints driver then twists each of the four flux tubes.
The fastest-rotating strand becomes kink-unstable after 1000 s and fragments into a turbulent structure of thinner strands. The initial helical current sheet fragments into many narrow current sheets. 
\textcolor{black}{The current sheets locally exceed the current density threshold for dissipation by anomalous resistivity ($J_{cr}$ in Eq. \ref{Eq:anomalus_diffusivity}}), and there the magnetic field reconnects impulsively. 
\textcolor{black}{Component-reconnection events occur intermittently across the guide magnetic field ($B_z$), producing localized bursts of heating and plasma acceleration. The resulting perturbations propagate across the field lines, interacting with neighboring, still-stable flux tubes. This coupling triggers further small-scale energy releases, collectively forming a nanoflare storm.
Here we investigate what happens afterwards, with sustained winding of the field lines, if and when a steady state is reached.}

Figure \ref{Fig:paper_1_3D_view} shows a 3D rendering of the simulation box at three different times, at the onset ($t = -1000\,\mathrm{s}$) and at the peak ($t = -740\,\mathrm{s}$) of the MHD avalanche, and after a relatively long time ($t = 1400\,\mathrm{s}$). The four flux tubes, subject to continuous, footpoints twisting, are emphasized with field lines of different colors. In particular, in the top row we show the atmospheric stratification of plasma density from one foot‑point to the midpoint of the tube (corresponding to the loop apex in a curved geometry). The second and third rows display, respectively, the magnitude of the current density and the temperature distribution, both restricted to coronal heights \textcolor{black}{($|z| < 2\,\mathrm{Mm}$)}, where $T > 10^{6}\,\mathrm{K}$. At $t = -1000\,\mathrm{s}$ the loop bundles are already highly twisted, the onset of the kink-instability is in the “yellow” strand, with a strong current and high-temperature shell enveloping the flux tube. The corona is still very tenuous ($n \sim 10^8\,\mathrm{cm}^{-3}$) and cold ($T \sim 0.8\,\mathrm{MK}$).
At $t = -740\,\mathrm{s}$ the instability has propagated across the neighbouring flux tubes. Reconnection occurs and the magnetic bundles lose their identity, as emphasized by the dispersed field lines. Chromospheric evaporation driven by the heating release starts to fill the corona with denser plasma (up to $10^9\,\mathrm{cm}^{-3}$), large current sheets ($\sim 2000\,\mathrm{StC}\,\mathrm{cm}^{-2}\,\mathrm{s}^{-1}$ during the strands disruption), and high temperatures  (peaks of $10\,\mathrm{MK}$) spread throughout the coronal part of the box.
At $t=1400\,\mathrm{s}$ the coronal magnetic field has irreversibly fragmented into many bundles of field lines continuously reconnecting with each other. Indeed, with the driving progressing through time, field lines follow convoluted paths along the coronal loop length (magnetic braiding) corresponding to high magnetic stresses. Therefore, many narrow current sheets form in the corona and impulsively heat the plasma with peaks in the temperature of $\sim 4\,\mathrm{MK}$ and an averaged value of $\sim 1.5\,\mathrm{MK}$. The atmosphere reaches a statistical heat balance with continuous footpoints twisting building up excess of magnetic energy, that is promptly released by DC heating events of nanoflare size.

Figure \ref{Fig:paper_1_time_evolution} summarizes the evolution of the simulation box from the initial, pre-avalanche twisting phase (phase 1., $t < -1000\,\mathrm{s}$), through the onset and rapid development of the instability (phase 2., $-1000 \le t < 0\,\mathrm{s}$) and the following post-avalanche phase (phase 3., $t > 0\,\mathrm{s}$) for about $1\,\mathrm{hr}$, which is longer than any typical timescale, e.g., sound crossing time, Alfvén time, thermal conduction time, radiative cooling time \citep{reale2007diagnostics}. 

In phase 1., temperature, density, and internal energy increase slowly and smoothly with time as the plasma is compressed in the twisted flux tube. The maximum velocity \textcolor{black}{and the total kinetic energy} (panels B  and G in Figure 2) settles to a steady value, below $100\,\mathrm{km}\,\mathrm{s}^{-1}$, due to impulsive reconnection events that release magnetic tension and accelerate the plasma \citep{2025A&A...695A..40C}. The maximum current (panel C) builds up linearly with time (with small spikes due to propagating waves) while the magnetic energy (panel E) shows a quadratic increase with time.
Phase 2. is characterized by sudden sharp peaks of temperature (above $10\,\mathrm{MK}$, panel A), velocity (above $300\,\mathrm{km}\,\mathrm{s}^{-1}$), electric currents, \textcolor{black}{and kinetic energy}, representing the disruption of the two couples of twisted strands at different times. The accumulated magnetic energy (panel E) drops dramatically by $50\%$ in a few hundred seconds while the internal energy (panel F) increases by a similar amount. In particular, the conversion of $\sim 10^{27}\,\mathrm{erg}$ into heat, together with the registered temperature peak of $\sim 10\,\mathrm{MK}$, matches with the observation of microflares, as those described by \citet{testa2020coronal}.
The plasma density (panel D) starts to increase gradually by chromospheric evaporation during phase 2. and reaches its peak at $t \sim 0\,\mathrm{s}$, after $\sim 1000\,\mathrm{s}$ from the strongest heat pulses, accordingly to the typical flaring loop timescales \citep{reale2007diagnostics, reale2014coronal}.
After this density peak of about $4 \times 10^{8}$ cm$^{-3}$, the density, as well as the magnetic energy and the internal energy, all settle to a rather steady value, much higher than the initial ones, with smooth oscillations with a period of about 2000 s (phase 3.).
Also, the maximum temperature, velocity, current density, \textcolor{black}{and kinetic energy} settle around constant values, much lower than the initial peaks, and high-frequency rapid fluctuations, all corresponding to continuous, randomly distributed, reconnection episodes. \textcolor{black}{Variations in the total energy (magnetic + internal + kinetic; red and dashed lines in panels E. and F.) are stronger during phase 1. when the magnetic strands load excess magnetic energy, and in the first $\sim 2000\,\mathrm{s}$ after the avalanche, when the thermal energy converted during the avalanche is radiated away. Kinetic energy is always negligible compared to the other two contributions.}

\subsection{Reconnection signatures}
\label{sec:reconnection_signatures}

One of the main goals of this study is to explore the reconnection details when the system has reached a steady state, in a rather dense and chaotic environment. We now focus on the evolution of a single field line subject to reconnection.

Figure \ref{Fig:paper_1_field_line_evolution} shows features of the temporal evolution of a selected field line. The line is retrieved by using a fourth-order
Runge-Kutta scheme, and starting from a point in the lower boundary. We follow the motion of the starting point according to the footpoints driver and reconstruct the field line from each position during the motion. The time-distance maps show, from top to bottom (left), the position of the field line in the x-direction, the temperature, and the ohmic heating. The motion of one (the uppermost) of the field line footpoints is shown on the right.
The three quantities show a behaviour alternating smooth evolution with rapid changes corresponding to sporadic and localized heating events and reconnection.

The three quantities all follow a similar, irregular behaviour. For extended periods, the field line drifts only slightly as the boundary driving slowly twists its footpoints; the temperature hovers near the coronal background of about $1\,\mathrm{MK}$, and the Ohmic‑heating sits close to zero. Every so often, however, the curve shows sudden sideways jumps of megameter size. These jumps coincide with equally sharp spikes in $J^{2}/\sigma$ and brief temperature rises of a few million kelvin degrees. 

For instance, the rapid variations (change in color) in time, around $t \sim 2500\,\mathrm{s}$, indicate a reconnection event (also visible as a long track in a small time in the right plot) occurring close to the loop apex ($z = 0\,\mathrm{Mm}$), as also marked by the arrows and circle. The field line breaks around z = 0 Mm and the lower half joins another line. In the figure, the upper half of the line suddenly jumps into another footpoint.
The second and third rows report rapid variation of temperature and strong and localized electric currents when reconnection occurs. 
Specifically, the reconnection leads to a significant heating release (circle) and temperature increase (arrow) from 1 MK to about 3 MK. In the right‑hand panels, regions of elevated temperature and enhanced heating rate coincide with the “upper‑footpoint jumps”, namely the abrupt connectivity changes produced by magnetic reconnection.
After the heating event, the flux tube cools down again, below 1 MK, due to radiation and thermal conduction.

\textcolor{black}{For typical post-heating coronal conditions in our simulation 
($L = 50~\mathrm{Mm}$, $T_max = 2.5~\mathrm{MK}$, $n = 10^{9}~\mathrm{cm^{-3}}$), 
the characteristic cooling times indicate that thermal conduction dominates 
the early decay of the heated plasma. Using the Spitzer conductivity 
$\kappa_{0} = 10^{-6}~\mathrm{erg\,s^{-1}\,cm^{-1}\,K^{-7/2}}$, 
the conductive cooling time is 
$\tau_{\mathrm{cond}} \simeq 3n k_{\mathrm{B}}L^{2}/(\kappa_{0}T^{5/2}) 
\approx 2\times10^{3}\,\mathrm{s}$. 
In contrast, the optically thin radiative cooling time, 
$\tau_{\mathrm{rad}} \simeq 3k_{\mathrm{B}}T/[n\Lambda(T)]$, 
with $\Lambda(T)\approx10^{-22}~\mathrm{erg\,cm^{3}\,s^{-1}}$ at 
$2.5~\mathrm{MK}$, yields 
$\tau_{\mathrm{rad}}\approx10^{4}\,\mathrm{s}$. 
Thus, conductive losses efficiently remove energy along the field lines 
shortly after the heating event, while radiative cooling becomes relevant 
during the later, lower-temperature phase. 
A minor contribution from adiabatic expansion may further assist the 
temperature decrease, consistent with previous analyses of coronal loop 
cooling timescales 
\citep{Hermans2021A&A655A36, Sen2024A&A688A64, Sen2025A&A699A106}.}

Fig. \ref{Fig:paper_1_physics_properties} shows maps at time $\sim 3000\,\mathrm{s}$ of the coronal flux tube during the impulsive dissipation of two current sheets. The map of the maximum ohmic heating rate (left panel) shows two main bundles of current sheets, where energy release is stronger than the surroundings and confined in a limited range of height around $z \sim 10\,\mathrm{Mm}$. This corresponds to two strips of higher temperature (middle panel), up to about $2.5\,\mathrm{MK}$. As expected, little direct evidence of the heating release can be found in the density map (right panel). There is a hint of evaporation around and above the transition region (arrow), at the footpoint corresponding to one of the strongest current sheets.

\subsection{Forward modelling}
\label{sec:forward_modelling}

We have synthesized observable quantities from the simulation as described in \citet{cozzo2024coronal}.  Instrument temperature response functions are calculated using CHIANTI 10 \citep{del2021chianti} considering ionization equilibrium, coronal element abundances \citep{feldman1992elemental}, constant electron density ($10^9\,\mathrm{cm}^{-3}$) and no absorption. \textcolor{black}{The broad filter EUV. bands were used in AIA synthetic imaging.}  We also accounted for AIA instrumental point spread functions (PSFs) \citep{poduval2013point}, while for MUSE PSFs we considered a Gaussian with FWHM of $0.45"$. Emission was then re-binned at instruments' pixel size (0.5" for AIA, 0.167" for MUSE).

In Fig. \ref{Fig:paper_1_AIA_maps}, we present synthetic AIA emission maps at the same time as in Fig. \ref{Fig:paper_1_physics_properties}. Each panel displays a side view of the intensity distribution integrated along the line of sight in the six EUV channels \citep{boerner2012initial}.  In all channels except for 131 \AA\ and 335 \AA\, we see diffuse emission along the entire coronal section of the flux tubes. In the 193 \AA\ and 211 \AA\ we see bright filamentary structures at the location of the hot plasma filaments shown in Fig.~\ref{Fig:paper_1_physics_properties}. In the 94 \AA\ and 171 \AA\ channels, we see bright spots just above the TR corresponding to the footpoints of reconnected field lines. While the emission in the 193 \AA\ and 211 \AA\ channels is due to the temperature enhancement in hot strands, the bursts observed at 94 \AA\ and 171 \AA\ are caused by the density increase, as chromospheric plasma expands in response to the deposition of heat in the lower atmosphere. 

Movie 1 shows the evolution of the emission in the six AIA channels. The initially hot ($\sim 2.5\,\mathrm{MK}$) coronal flux tube is bright at 193 \AA\; after 400 s it cools down to $\sim 1\,\mathrm{MK}$ fainting in the 193 \AA\ filter and brightening up at 94 \AA\ and 171 \AA\. The emission in those channels is higher in the lower corona and relatively uniform in the cross-field direction. Between $t=1000\,\mathrm{s}$ and $t=2000\,\mathrm{s}$ the atmosphere is faint in all the AIA channels: this is the period where both plasma density and internal energy show a dip in the time profiles shown in Figure \ref{Fig:paper_1_time_evolution}. A secondary, post-avalanche, collective energy release is registered starting from $t=2000\,\mathrm{s}$. This is the outcome of several small-scale heating events visible as narrow strips both in the 193 \AA\ and 211 \AA\ channels.  

Fig. \ref{Fig:paper_1_MUSE_maps} shows a side-view of the flux tube emission, Doppler shift along the line of sight, and non-thermal line broadening in the MUSE \feix\ line at 171 \AA\ and \fexv\ at 284 \AA\ at the same snapshot as in Figs. \ref{Fig:paper_1_physics_properties} and \ref{Fig:paper_1_AIA_maps}. The non-thermal velocity and width in each channel are computed as the first and second momenta of the velocity distribution, weighted for the emissivities, as described in \citet{cozzo2024coronal} and \citet{de2022probing}. The MUSE \feix\ emission closely resembles that of the AIA 171 \AA\ channel, including similarly brightened footpoints as indicated by the arrows. The Doppler shifts map shows an alternating pattern of red- and blue-shifted velocity, at magnitudes up to about 20 km$\,$s$^{-1}$, organized into elongated, thin strands. These slow Alfvénic and tortional motions captured by the cold channel do not show strong correlation with any impulsive heating event, although their oscillating behaviour is likely powered both by the footpoints driver and the residual dynamics released in the corona during reconnection. Likewise, the line broadening exhibits a filamentary structure punctuated by localized regions with significant broadening (up to $\sim 25\,\mathrm{km}\,\mathrm{s}^{-1}$). Conversely, emission in the MUSE 284\,\AA\ line is bright only within two very narrow strips, clearly traceable back to the locations of heating deposition described in Fig. \ref{Fig:paper_1_physics_properties}, and correlating with the bright footpoint spots observed in the AIA 193 \AA\ and 211 \AA\ channels. The Doppler-shift maps in the MUSE 284 \AA\ line distinctly capture bidirectional outflows from the locations of peak brightness. Additionally, the line broadening appears the strongest in the core of the emission regions, as compared to their surroundings.

Movie 2, shows the evolution of emission, Doppler velocity, and non-thermal line broadening in the MUSE \feix\ and \fexv\ channels. \feix\ emission evolves analogously to the AIA 171\AA\ filter, although the enhanced resolution allows the detection of small-scale, cross-field structuring. Doppler velocity shows alternation of blue and red shifted regions while the non-thermal width appears filamented, with sub-arcsec structuring across the guide field. \fexv\ brightens up only sporadically within fine strips, always accompanied by strong enhancement in the non-thermal line broadening and Doppler shifts marking bidirectional outflows departing from the bright spots.

\subsection{Effects of magnetic diffusivity on heating}
\label{sec:effects_diffusivity}

To investigate the role of dissipation, we performed three additional simulations.
We considered a guide magnetic field of $10\,\mathrm{G}$, and we varied the magnetic diffusivity ($\eta$) between $10^{11}$ and $10^{13}\,\mathrm{cm}^{2}\,\mathrm{s}^{-1}$, and the smallest spatial cell width ($\Delta$) between $25\,\mathrm{km}$ and $50\,\mathrm{km}$, as summarized in table \ref{table_1}. The current density threshold is chosen to scale with the intensity of the magnetic field ($150\,\mathrm{StC}\,\mathrm{cm}^2$ for $B = 10\,\mathrm{G}$, $300\,\mathrm{StC}\,\mathrm{cm}^2$ for $B = 20\,\mathrm{G}$) in order to generate current sheets of similar thickness. 
\textcolor{black}{For the adopted plasma parameters (assuming $n = 10^{9},\mathrm{cm^{-3}}$), the resulting Lundquist numbers span from $(S \sim 10^{1})$ in run 3 to $(S \sim 10^{3}!-!10^{4})$ in the other cases (Table \ref{table_1}). These values indicate that magnetic reconnection proceeds in a moderately resistive regime. Runs 1, 2, and 4, with $(S \gtrsim 10^{3})$, lie close to the threshold $((S \sim 10^{4}-10^{5}))$ at which the plasmoid instability can begin to fragment current sheets into multiple secondary islands, leading to bursty, turbulent reconnection. In contrast, run 3 (with the higher resistivity and $(S \sim 10^{1})$) falls well below this threshold and is expected to exhibit laminar, Sweet–Parker–like reconnection.}

\begin{table}
\caption{Simulation parameters and corresponding Lundquist numbers.}
\centering
\begin{tabular}{lcccccc}
\hline\hline
Sim. & B & $\eta$ & $J_{\mathrm{cr}}$ & $\Delta$ & $S$ \\
 & [G] & [$\mathrm{cm}^{2}\,\mathrm{s}^{-1}$] & [StC cm$^{-2}$ s$^{-1}$] & [km] &  \\
\hline
1 & 20 & $10^{11}$ & 300 & 25 & $3.5\times10^{3}$ \\
2 & 10 & $10^{11}$ & 150 & 25 & $1.8\times10^{3}$ \\
3 & 10 & $10^{13}$ & 150 & 25 & $1.8\times10^{1}$ \\
4 & 10 & $10^{11}$ & 150 & 50 & $3.5\times10^{3}$ \\
\hline
\end{tabular}
\label{table_1}
\end{table}

\begin{figure*}[h!]
   \centering
\includegraphics[width=\hsize]{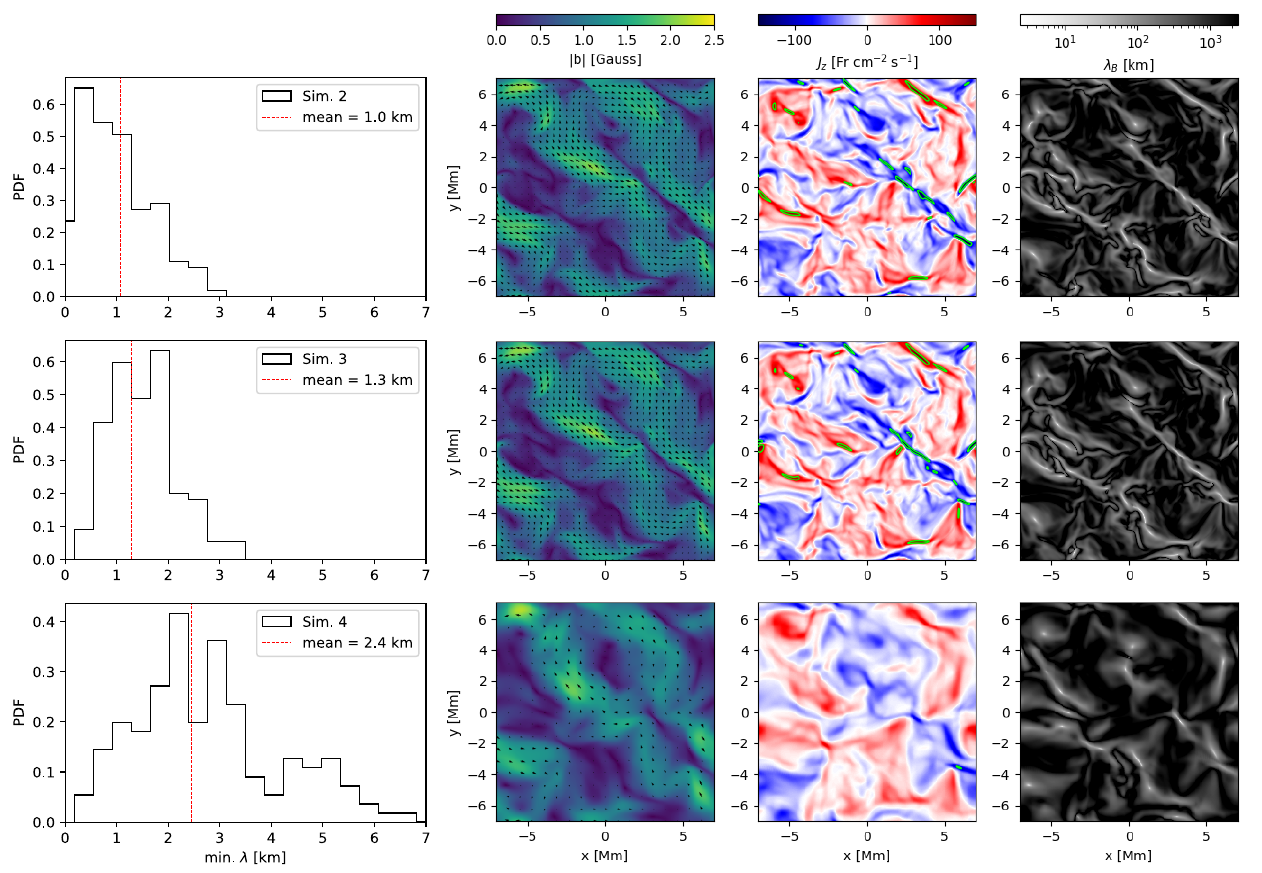}
    \caption{Effects of magnetic diffusion on current sheets. {\it Left:} The histograms of the distribution of the minimum current sheets thickness ($\lambda_B = \frac{c}{4 \pi}|b|/|J_z|$) for simulations 2, 3, and 4, listed in Table \ref{table_1}, with different resolution ($\Delta_{min} = 25,\,50\,\mathrm{km}$), and magnetic diffusion ($\eta = 10^{11},\,10^{13}\,\mathrm{cm}^{2}\,s^{-1}$). 
  {\it Right:} The maps of the horizontal magnetic field (2nd column), vertical current density (3rd column), and $\lambda_B$ (4th column) distribution on a horizontal cut at the mid-plane, $\sim 1500\,\mathrm{s}$ after the onset of the avalanche, for each of the three simulations.} \label{Fig:paper_1_parameter_sampling_1}
\end{figure*}

In Fig.\ref{Fig:paper_1_parameter_sampling_1}, we explore how magnetic diffusion influences the formation, morphology, and strength of the current sheets generated by the magnetic field shear, which subsequently dissipate into heat due to anomalous resistivity. Specifically, the histograms in the first column show the distribution of the minimum current sheet thickness ($\lambda_B$) at each temporal output, for runs 2-to-4. The current sheets develop mostly along the strong guide field in the $\hat{z}$ direction. We estimate the characteristic thickness of these thin layers by the ratio between the magnitude of the transverse (``component'') magnetic field, $|b| = \sqrt{b_x^2 + b_y^2}$ (lying in the $\hat{x}$–$\hat{y}$ plane, second column of Fig. \ref{Fig:paper_1_parameter_sampling_1}), and the magnitude of the current density along $\hat{z}$ ($J_z$, third column); 
\begin{equation}
\lambda_B = \frac{c}{4 \pi}\frac{|b|}{|J_z|}.
\end{equation}
An example of the distribution of \textcolor{black}{$\lambda_B$} at the mid-plane of the box (loop apex) is shown in the fourth column.
Simulation 2 (first row in Figure \ref{Fig:paper_1_parameter_sampling_1}) is the least current-dissipative because it has both the smallest magnetic resistivity and the highest resolution, as compared with runs 3 (higher $\eta$) and 4 (lower resolution). Increasing the magnetic diffusivity in run 3 (second row in Figure \ref{Fig:paper_1_parameter_sampling_1}) results in a larger average thickness of the current sheets, with the effect being significantly more pronounced when the numerical diffusivity is higher due to the coarser spatial grid, as in run 4 (third row in Figure \ref{Fig:paper_1_parameter_sampling_1}).

\begin{figure*}[h!]
   \centering
\includegraphics[width=\hsize]{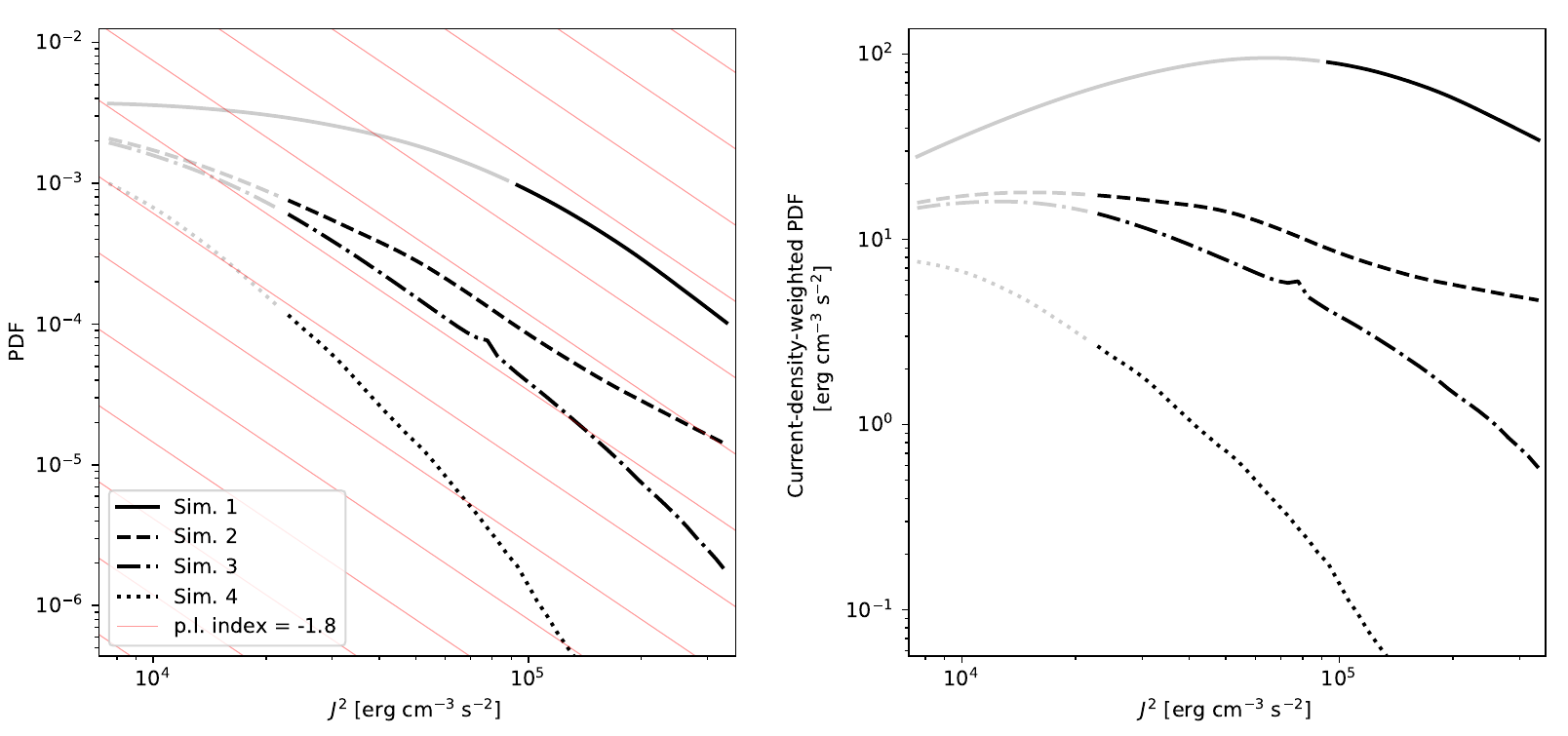}
  \caption{Effects of magnetic diffusion on the heating. On the left panel we show the time‐averaged probability density functions (PDFs) of the current density square
  (proportional to the volumetric Ohmic heating rate) 
  for 4 numerical runs with magnetic field strength 10~G and 20~G, minimum spacing $\Delta=25$~km and 50~km, and magnetic diffusivity $\eta=10^{11}$ and $10^{13}$~cm$^{2}$\,s$^{-1}$). 
  Specifically, opaque regions mark heating from anomalous dissipation ($J > J_{\mathrm{cr}}$), while lighter regions still show the $J^{2}$ distribution, which the code does not explicitly dissipate (but which may be subject to numerical diffusion). The red lines mark a reference power law with index $-1.8$, illustrating the canonical nanoflare‑size distribution expected for impulsive heating events in the solar corona.
  Similarly, on the right, we show the 
  current-density‑weighted PDFs.}
\label{Fig:paper_1_parameter_sampling_2}
\end{figure*}

We now examine how the intensity distribution of the heating depends on magnetic field strength, magnetic diffusivity, and spatial resolution. The left panel in figure \ref{Fig:paper_1_parameter_sampling_2} shows the time-averaged probability density functions (PDFs) of the square of the current density (proportional to the volumetric Ohmic-heating rate $E = J^{2}/\sigma$), i.e., the fraction of the volume $V(J^2)$ where the local heating rate falls within each (logarithmic) current density bin. The curves in the right panel show the same quantity multiplied by the respective current density square, as an estimate of the ohmic energy carried by each current bin. 
Transparent regions mark sub-critical currents ($J<J_{\mathrm{cr}}$) where no heating from anomalous magnetic diffusivity is occurring. The left PDFs show that the occurrence frequency in the heating regime closely follows a power law $\propto E^{-1.8}$ expected for a nanoflare distribution \citep{hudson2010observations}.
The right PDFs show a rather flat energy contribution across the inertial range, so that no single scale dominates the total heating budget. Deviations at the lowest energies are set by the numerical diffusion floor, while the high‑energy cut‑off reflects the finite magnetic free‑energy reservoir. Together, the PDFs show that threshold‑triggered dissipation not only localizes heating to strongly current‑bearing strands but also reproduces the expected statistical signatures (both in event counts and in energetic impact) of nanoflare‑driven coronal heating.
Generally, electric currents tend to accumulate until reaching the threshold for anomalous dissipation, as illustrated by the current-density-weighted PDFs (right).
As expected, a stronger magnetic field results in more intense heating (Sim. 1 vs. Sim. 2). Conversely, a lower spatial resolution inhibits the build-up of currents, significantly reducing the amount of heating released. Higher magnetic diffusivity, as in Sim. 3, dissipates magnetic stresses more efficiently, thereby suppressing the formation of intense current sheets, as shown by the increasing distance between the dashed (Sim. 2) and dot-dashed (Sim. 3) curves at high values of $J^2$.

\section{Discussion and Conclusions}
\label{sec:conclusions}
In \cite{cozzo2023coronal} we have explored how kink-unstable flux tubes fragment into chaotic current sheets where magnetic reconnection triggers impulsive heating release, through anomalous resistivity, in the presence of a stratified atmosphere from the chromosphere to the corona. The faster-rotating flux tube fragments first and propagates the instability to the other slower one, thus describing an avalanche process which might explain the ignition of a larger scale coronal loop. The initial heat pulses in a relatively empty atmosphere lead to strong local temperature peaks (up to about 10 MK), before evaporation fronts bring up dense plasma to fill the corona. It is then interesting to investigate what happens on a longer time scale with continuous driving of the twisting, in particular if and what steady-state conditions are reached, and then what is the dependence of these conditions on some crucial parameters. Specific questions are: does the magnetic field remain capable of storing the energy continuously provided by footpoint motions, ultimately triggering nanoflare storms, or is the accumulated energy quickly released through frequent but less energetic nanoflares \citep{rappazzo2013field}?

To address these questions, the stratified atmosphere  \citep{cozzo2023coronal} is important because it allows us to study accurately the plasma response  to the impulsive heating in the corona and chromosphere, and for forward-modeling observations of existing (e.g., AIA/SDO and IRIS) and forthcoming (MUSE) observations \citep{cozzo2023coronal}.

Also in this work, we consider three-dimensional MHD simulations of an MHD avalanche occurring within a kink-unstable, multi-threaded coronal loop system \citep[see also][]{2016mhhoodd, reid2018coronal, reid2020coronal, cozzo2023coronal}. At variance from the previous work, here we consider four, instead of two, thinner magnetic flux tubes. The critical amount of twist for kink-instability is therefore smaller \citep{hood1979equilibrium} and the four initial strands get fragmented in a shorter period of time.
As before, they are embedded in a stratified atmosphere, characterized by a corona at $1\,\mathrm{MK}$ anchored on both ends to a denser, cooler ($10^4\,\mathrm{K}$) isothermal chromosphere. Rotational motions at the loop footpoints progressively twist these flux tubes in an ambient magnetic field of strength $B_{\mathrm{bkg}}=20\,\mathrm{Gauss}$. A small anomalous magnetic resistivity ($10^{11}\,\mathrm{cm}^{2}\,\mathrm{s}^{-1}$  and $J_{\mathrm{cr}}$) facilitates
the build-up of strong currents in narrow current sheets, making the heating impulsive and localized.

As twisting continues, the flux tubes become kink-unstable and rapidly fragment into a chaotic structure filled with thin current sheets, hosting small-scale impulsive reconnection events. In particular, the two couples of flux ropes merge by magnetic reconnection at different times \citep{browning1986heating}. These events locally and impulsively heat the plasma to peak temperatures exceeding $10\,\mathrm{MK}$ as soon as the instability propagates. The widespread intense nanoflares also trigger chromospheric evaporation, thereby enhancing loop brightness in the EUV band \citep{cozzo2024coronal}. 

We follow the evolution of the system for over 1 hour after the avalanche is triggered, longer than typical cooling times.
In our model, continuous driving motions at the loop footpoints \citep{reid2018coronal, reale20163d} consistently inject magnetic energy into the corona. Consequently, the pre-fragmented magnetic field configuration naturally facilitates the formation of additional current sheets, and sustained Ohmic dissipation through anomalous resistivity. Specifically, the coronal loop remains endlessly braided, undergoing intermittent episodes of resistive relaxation. Its tangled magnetic field triggers spontaneous reconnection, producing a cascade of small-scale nanoflares \citep{pontin2011dynamics}. Eventually, the system gets to a balance between radiative losses, thermal conduction, and continuous coronal DC heating, and a statistical steady-state equilibrium is established \citep{reid2020coronal}. 
Following each reconnection event, the field attempts to relax toward a lower-energy configuration, yet the continuous application of driving velocities, leading to sustained injection of energy from boundary motions, prevents the magnetic system from approaching the minimal-energy state predicted by classical relaxation theory \citep{taylor1974relaxation}. 
Temperature and current density are highly structured at sub-observational scales, making the loop plasma multithermal, with dynamic current strands scattered throughout. Consequently, coronal heating is found to be highly intermittent, with most of the loop cooling at any given time \citep{dahlburg2016observational}.

Previous studies propose that current sheets can form with sufficiently tangled magnetic fields driven by complex boundary motions \citep{hendrix1996magnetohydrodynamic, ng2012high}. 
In this scenario, the relatively simple vortex motions imposed at the boundaries lead to initially distinct and coherent magnetic flux tubes. Nevertheless, the onset of instability leads to significant changes in magnetic connectivity. 
As a consequence, magnetic field lines become linked to different vortices, and the ongoing twisting boundary motions begin to braid the field lines into an increasingly complex configuration.
Therefore, current sheets spontaneously emerge from a highly ordered initial state, with sustained footpoints motions gradually braiding the magnetic field, a phenomenon frequently observed in similarly driven simulations \citep{reid2018coronal, reid2020coronal, wilmot2015overview}.

Figure \ref{Fig:paper_1_time_evolution} clearly shows the initial break up that determines peaks of temperature around 10 MK, in agreement with \citet{cozzo2023coronal}. Afterwards, both the gradual filling of the tubes by chromospheric evaporation, that determines an increase of the plasma heat capacity, and the continuous energy release, that inhibits large energy accumulation for strong events, prevent the loop plasma from reaching the initial high temperature peaks; instead, the loop shows a continuous sequence of smaller peaks.
These subsequent heating events exhibit neither clear periodicity nor characteristic magnitude \citep{reid2020coronal}.
Average temperature and density achieved by the system at regime are comparable with the results found by \cite{howson2022effects} with a guide field strength of $20\,\mathrm{G}$.

The synchronized signatures of current density and temperature mark clear reconnection pulses (Figure \ref{Fig:paper_1_field_line_evolution}), field lines reconnect through narrow current sheets, their magnetic connections rearrange almost instantaneously, and the stored magnetic energy is dissipated locally as heat. Once the burst is over, the plasma cools by conduction and radiation, and the line settles back into its slow drift until the next trigger. This alternation of gradual build‑up and rapid release is the textbook pattern expected for nanoflare‑style heating of coronal loops.

\cite{2025A&A...695A..40C} demonstrated that the detection of localized heating events is facilitated by examining the emission of narrow, hot spectral lines (e.g., MUSE \fexix\  108 \AA), as compared to broad-band AIA channels such as 94 \AA, in which emission from hot lines is blended with cooler background emission (e.g., \citealt{Reale2011,Foster2011,Testa2012,Testa2012b,DelZanna2013}). Though in a different regime, a similar conclusion can be drawn from the present work by comparing the emission maps of the MUSE \fexv\  284 \AA\ line with synthetic emission images derived from the AIA 193 \AA\ and 211 \AA\ channels.

As illustrated in Fig. \ref{Fig:paper_1_MUSE_maps}, since in the regime of our first simulation here the average temperature settles to approximately $1\,\mathrm{MK}$, punctuated by temperature spikes around $2.5\,\mathrm{MK}$ in transient and bright strands \citep{breu2022solar},  the \feix\  171 \AA\ line emission is dominated by background emission from cooling plasma (with bright spots at the footpoints), whereas the \fexv\ 284\AA\ line emission clearly highlights localized heating events. 

In this paper, we also explored various conditions of resistivity and resolution (see table \ref{table_1}). Smaller resistivity and higher grid resolution lead to strong currents (Fig. \ref{Fig:paper_1_parameter_sampling_2}) in finer sheets (Fig. \ref{Fig:paper_1_parameter_sampling_1}), as reduced (physical and numerical) dissipation allows stresses to build up to higher levels \citep{rappazzo2007coronal, rappazzo2008nonlinear}.  Stronger currents lead in turn to more intense heating events (Fig. \ref{Fig:paper_1_parameter_sampling_2}), described with a frequency-energy power law with index $\sim -1.8$, in agreement with  the ``flare syndrome theory'' \citep{hudson2010observations} and supported by numerous observations at higher energies \citep{drake1971characteristics, dennis1985solar, crosby1993frequency}. 

In conclusion, in this work we verified that, after an impulsive transient due to the onset of the kink instability, on long time scales a continuous footpoint driver of a multiple flux tube system leads to an almost steady-state regime. The frequent power-law distributed nanoflares heat the plasma to lower temperatures than in the transient and keep the system at a 1 MK regime with the selected background magnetic field, with interesting implications on the observables. We have also confirmed that the diffusion and resolution parameters are important to determine the intensity of the bursts. Future investigations will explore different regimes, for instance those of active regions.

\begin{acknowledgments}
      GC and PT were supported by contract 4105785828 (MUSE) to the Smithsonian Astrophysical Observatory, and by NASA grant 80NSSC21K1684.
      PP, and FR acknowledge support from ASI/INAF agreement n. 2022-29-HH.0. 
      This work made use of the HPC system MEUSA, part of the Sistema Computazionale per l'Astrofisica Numerica (SCAN) of INAF-Osservatorio Astronomico di Palermo.
      The simulations have been run on the Pleiades cluster through the computing projects s1061, s2601, from the High-End Computing (HEC) division of NASA.
      We acknowledge the CINECA award under the ISCRA initiative, for the availability of high performance computing resources and support.
      BDP and JMS were supported by NASA contract 80GSFC21C0011 (MUSE) and NNG09FA40C (IRIS).
\end{acknowledgments}

\software{PLUTO \citep{mignone2007pluto}}

\appendix
\section{Radiative losses}
\label{sec:appendix}
\begin{figure}
    \centering
    \includegraphics[width=\hsize]{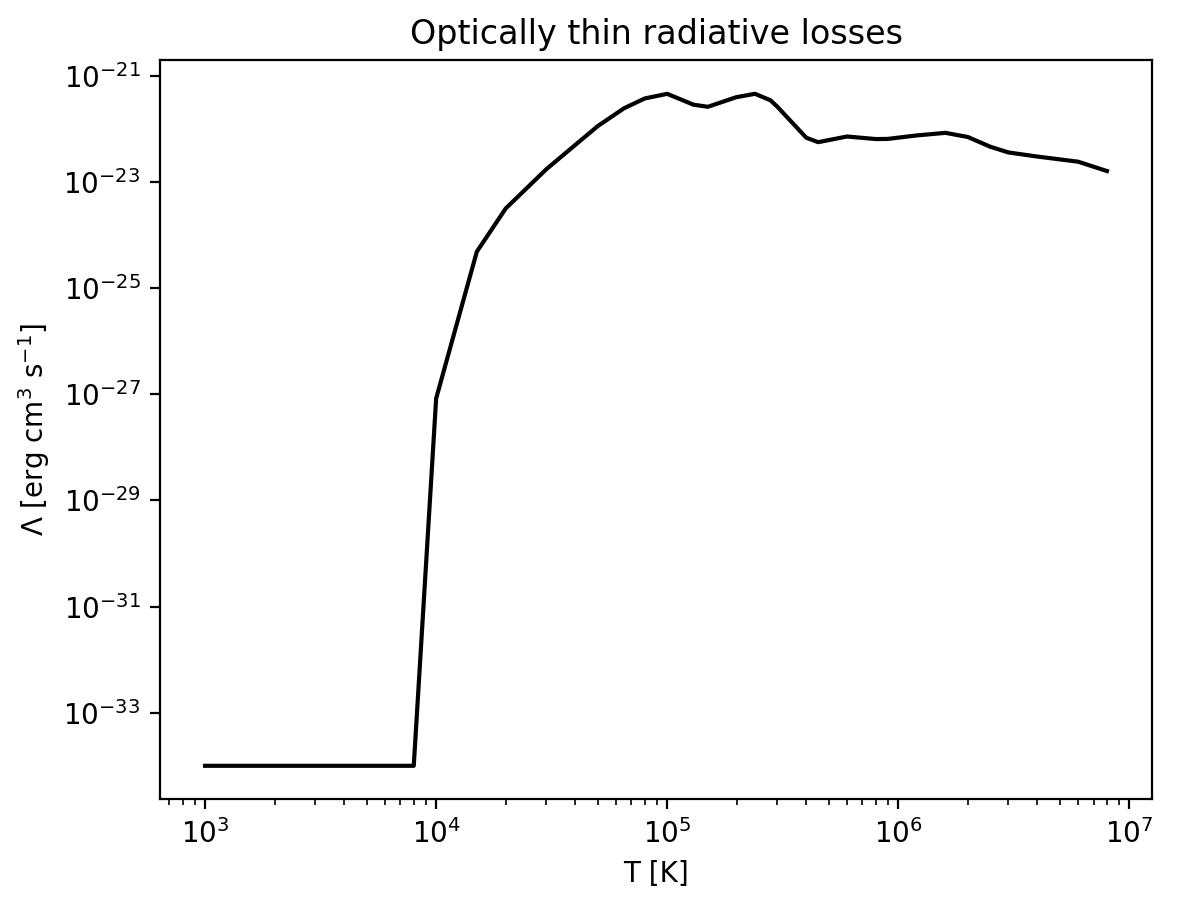}
    \caption{Function of optically thin radiative losses per unit emission measure, derived from the CHIANTI v. 7.0 database \citep{landi2013prominence}, assuming coronal element abundances \citep{widing1992element}.}
    \label{fig:Opticallythinlosses}
\end{figure}

Atmospheric gravitational stratification on the Sun is modulated by radiative and heat transfer. In the corona, collision de-excitation and photon absorption are negligible, generated photons directly escape. Optically thin radiative cooling is important in the corona with a rate per unit volume:
\begin{equation}
    Q_{\text{coro}} = n_{\mathrm{H}} n_{\mathrm{e}} \Lambda(T),
    \label{Eq:PLUTO_optically_thin_loss}
\end{equation}
where $\Lambda (T)$  is the plasma emissivity per unit emission measure (integrated over all wavelengths); $n_{\mathrm{H}}$ and $n_{\mathrm{e}}$  are the hydrogen and electron number density, respectively (assumed equal).
PLUTO code includes optically thin radiative losses in a fractional step formalism, which preserves the second-order time accuracy, since the advection and source steps are at least second-order accurate; the radiative loss values ($\Lambda (T)$) are computed at the temperature of interest using a table lookup/interpolation method. 
In particular, we considered the radiative losses from optically thin plasma per unit emission measure (shown in the plot of Fig. \ref{fig:Opticallythinlosses}), derived from the CHIANTI v. 7.0 database \citep[e.g.,][]{landi2013prominence} assuming coronal element abundances \citep{widing1992element}.

\bibliography{sample701}{}

@ARTICLE{DelZanna2013,
       author = {{Del Zanna}, G.},
        title = "{The multi-thermal emission in solar active regions}",
      journal = {\aap},
         year = 2013,
        month = oct,
       volume = {558},
          eid = {A73},
        pages = {A73},
          doi = {10.1051/0004-6361/201321653},
       adsurl = {https://ui.adsabs.harvard.edu/abs/2013A&A...558A..73D}
}

@ARTICLE{Reale2011,
       author = {{Reale}, Fabio and {Guarrasi}, Massimiliano and {Testa}, Paola and {DeLuca}, Edward E. and {Peres}, Giovanni and {Golub}, Leon},
        title = "{Solar Dynamics Observatory Discovers Thin High Temperature Strands in Coronal Active Regions}",
      journal = {\apjl},
         year = 2011,
        month = jul,
       volume = {736},
       number = {1},
          eid = {L16},
        pages = {L16},
          doi = {10.1088/2041-8205/736/1/L16},
archivePrefix = {arXiv},
       eprint = {1106.1591},
 primaryClass = {astro-ph.SR},
       adsurl = {https://ui.adsabs.harvard.edu/abs/2011ApJ...736L..16R}
}

@ARTICLE{Testa2012b,
       author = {{Testa}, Paola and {Reale}, Fabio},
        title = "{Hinode/EIS Spectroscopic Validation of Very Hot Plasma Imaged with the Solar Dynamics Observatory in Non-flaring Active Region Cores}",
      journal = {\apjl},
         year = 2012,
        month = may,
       volume = {750},
       number = {1},
          eid = {L10},
        pages = {L10},
          doi = {10.1088/2041-8205/750/1/L10},
archivePrefix = {arXiv},
       eprint = {1204.0041},
 primaryClass = {astro-ph.SR},
       adsurl = {https://ui.adsabs.harvard.edu/abs/2012ApJ...750L..10T}
}

@ARTICLE{Testa2012,
       author = {{Testa}, Paola and {Drake}, Jeremy J. and {Landi}, Enrico},
        title = "{Testing EUV/X-Ray Atomic Data for the Solar Dynamics Observatory}",
      journal = {\apj},
         year = 2012,
        month = feb,
       volume = {745},
       number = {2},
          eid = {111},
        pages = {111},
          doi = {10.1088/0004-637X/745/2/111},
archivePrefix = {arXiv},
       eprint = {1110.4611},
 primaryClass = {astro-ph.SR},
       adsurl = {https://ui.adsabs.harvard.edu/abs/2012ApJ...745..111T}
}

@ARTICLE{Foster2011,
       author = {{Foster}, Adam R. and {Testa}, Paola},
        title = "{Fe IX Calculations for the Solar Dynamics Observatory}",
      journal = {\apjl},
         year = 2011,
        month = oct,
       volume = {740},
       number = {2},
          eid = {L52},
        pages = {L52},
          doi = {10.1088/2041-8205/740/2/L52},
archivePrefix = {arXiv},
       eprint = {1107.0470},
 primaryClass = {astro-ph.SR},
       adsurl = {https://ui.adsabs.harvard.edu/abs/2011ApJ...740L..52F}
}

@ARTICLE{testa2015,
       author = {{Testa}, Paola and {Saar}, Steven H. and {Drake}, Jeremy J.},
        title = "{Stellar activity and coronal heating: an overview of recent results}",
      journal = {Philosophical Transactions of the Royal Society of London Series A},
         year = 2015,
        month = apr,
       volume = {373},
       number = {2042},
        pages = {20140259-20140259},
          doi = {10.1098/rsta.2014.0259},
archivePrefix = {arXiv},
       eprint = {1502.07401},
 primaryClass = {astro-ph.SR},
       adsurl = {https://ui.adsabs.harvard.edu/abs/2015RSPTA.37340259T}
}

@article{rappazzo2007coronal,
  title={Coronal heating, weak MHD turbulence, and scaling laws},
  author={Rappazzo, Antonio F and Velli, Marco and Einaudi, Giorgio and Dahlburg, RB},
  journal={The Astrophysical Journal},
  volume={657},
  number={1},
  pages={L47},
  year={2007},
  publisher={IOP Publishing}
}

@article{rappazzo2008nonlinear,
  title={Nonlinear dynamics of the Parker scenario for coronal heating},
  author={Rappazzo, AF and Velli, Marco and Einaudi, Giorgio and Dahlburg, RB},
  journal={The Astrophysical Journal},
  volume={677},
  number={2},
  pages={1348},
  year={2008},
  publisher={IOP Publishing}
}

@article{rappazzo2013field,
  title={Field lines twisting in a noisy corona: implications for energy storage and release, and initiation of solar eruptions},
  author={Rappazzo, AF and Velli, M and Einaudi, G},
  journal={The Astrophysical Journal},
  volume={771},
  number={2},
  pages={76},
  year={2013},
  publisher={IOP Publishing}
}

@article{dahlburg2016observational,
  title={Observational signatures of coronal loop heating and cooling driven by footpoint shuffling},
  author={Dahlburg, RB and Einaudi, G and Taylor, BD and Ugarte-Urra, I and Warren, HP and Rappazzo, AF and Velli, M},
  journal={The Astrophysical Journal},
  volume={817},
  number={1},
  pages={47},
  year={2016},
  publisher={IOP Publishing}
}

@article{breu2022solar,
  title={A solar coronal loop in a box: energy generation and heating},
  author={Breu, C and Peter, H and Cameron, R and Solanki, SK and Przybylski, D and Rempel, M and Chitta, LP},
  journal={Astronomy \& Astrophysics},
  volume={658},
  pages={A45},
  year={2022},
  publisher={EDP Sciences}
}

@article{vogler2005simulations,
  title={Simulations of magneto-convection in the solar photosphere-Equations, methods, and results of the MURaM code},
  author={V{\"o}gler, Alexander and Shelyag, Sergiy and Sch{\"u}ssler, Manfred and Cattaneo, Fausto and Emonet, Thierry and Linde, Timur},
  journal={Astronomy \& Astrophysics},
  volume={429},
  number={1},
  pages={335--351},
  year={2005},
  publisher={EDP Sciences}
}

@article{rempel2016extension,
  title={Extension of the MURaM radiative MHD code for coronal simulations},
  author={Rempel, Matthias},
  journal={The Astrophysical Journal},
  volume={834},
  number={1},
  pages={10},
  year={2016},
  publisher={IOP Publishing}
}

@article{pontin2011dynamics,
  title={Dynamics of braided coronal loops-II. Cascade to multiple small-scale reconnection events},
  author={Pontin, DI and Wilmot-Smith, AL and Hornig, G and Galsgaard, K},
  journal={Astronomy \& Astrophysics},
  volume={525},
  pages={A57},
  year={2011},
  publisher={EDP Sciences}
}

@article{tam2015coronal,
  title={Coronal heating in multiple magnetic threads},
  author={Tam, Kuan Vai and Hood, Alan William and Browning, PK and Cargill, PJ},
  journal={Astronomy \& Astrophysics},
  volume={580},
  pages={A122},
  year={2015},
  publisher={EDP Sciences}
}

@article{2016mhhoodd,
  title={An MHD avalanche in a multi-threaded coronal loop},
  author={Hood, Alan William and Cargill, PJ and Browning, PK and Tam, KV},
  journal={The Astrophysical Journal},
  volume={817},
  number={1},
  pages={5},
  year={2016},
  publisher={IOP Publishing}
}

@article{browning1986heating,
  title={Heating of coronal arcades by magnetic tearing turbulence, using the Taylor-Heyvaerts hypothesis},
  author={Browning, PK and Priest, ER},
  journal={Astronomy and Astrophysics},
  volume={159},
  pages={129--141},
  year={1986}
}

@article{gold1960origin,
  title={On the origin of solar flares},
  author={Gold, TRACEY and Hoyle, F},
  journal={Monthly Notices of the Royal Astronomical Society},
  volume={120},
  number={2},
  pages={89--105},
  year={1960},
  publisher={Oxford Academic}
}

@article{melrose1997solar,
  title={A solar flare model based on magnetic reconnection between current-carrying loops},
  author={Melrose, DB},
  journal={The Astrophysical Journal},
  volume={486},
  number={1},
  pages={521},
  year={1997},
  publisher={IOP Publishing}
}

@article{kondrashov1999three,
  title={Three-dimensional Magnetohydrodynamic Simulationsof the Interaction of Magnetic Flux Tubes},
  author={Kondrashov, D and Feynman, J and Liewer, PC and Ruzmaikin, A},
  journal={The Astrophysical Journal},
  volume={519},
  number={2},
  pages={884},
  year={1999},
  publisher={IOP Publishing}
}

@article{linton2001reconnection,
  title={Reconnection of twisted flux tubes as a function of contact angle},
  author={Linton, MG and Dahlburg, RB and Antiochos, SK},
  journal={The Astrophysical Journal},
  volume={553},
  number={2},
  pages={905},
  year={2001},
  publisher={IOP Publishing}
}

@article{kliem2014slow,
title = {Slow Rise and Partial Eruption of a Double-Decker Filament. II. A Double Flux Rope Model},
author = {Kliem, Bernhard and T{\"o}r{\"o}k, Tibor and Titov, Viacheslav S. and Lionello, Roberto and Linker, Jon A. and Liu, Rui and Liu, Chang and Wang, Haimin},
journal = {The Astrophysical Journal},
doi = {10.1088/0004-637X/792/2/107},
volume = {792},
number = {2},
pages = {107},
year = {2014},
publisher = {The American Astronomical Society},
}

@article{liu2020magnetic,
  title={Magnetic flux ropes in the solar corona: structure and evolution toward eruption},
  author={Liu, Rui},
  journal={Research in Astronomy and Astrophysics},
  volume={20},
  number={10},
  pages={165},
  year={2020},
  publisher={IOP Publishing}
}

@article{reid2018coronal,
  title={Coronal energy release by MHD avalanches: continuous driving},
  author={Reid, Jack and Hood, Alan William and Parnell, Clare Elizabeth and Browning, PK and Cargill, PJ},
  journal={Astronomy \& Astrophysics},
  volume={615},
  pages={A84},
  year={2018},
  publisher={EDP Sciences}
}

@article{reid2020coronal,
  title={Coronal energy release by MHD avalanches: Heating mechanisms},
  author={Reid, Jack and Cargill, PJ and Hood, Alan William and Parnell, Clare Elizabeth and Arber, Tony Deane},
  journal={Astronomy \& Astrophysics},
  volume={633},
  pages={A158},
  year={2020},
  publisher={EDP Sciences}
}

@article{hendrix1996magnetohydrodynamic,
  title={Magnetohydrodynamic turbulence and implications for solar coronal heating},
  author={Hendrix, DL and Van Hoven, G},
  journal={Astrophysical Journal v. 467, p. 887},
  volume={467},
  pages={887},
  year={1996}
}

@article{ng2012high,
  title={High-Lundquist number scaling in three-dimensional simulations of Parker's model of coronal heating},
  author={Ng, CS and Lin, L and Bhattacharjee, A},
  journal={The Astrophysical Journal},
  volume={747},
  number={2},
  pages={109},
  year={2012},
  publisher={IOP Publishing}
}

@article{wilmot2015overview,
  title={An overview of flux braiding experiments},
  author={Wilmot-Smith, AL},
  journal={Philosophical Transactions of the Royal Society A: Mathematical, Physical and Engineering Sciences},
  volume={373},
  number={2042},
  pages={20140265},
  year={2015},
  publisher={The Royal Society Publishing}
}

@article{taylor1974relaxation,
  title={Relaxation of toroidal plasma and generation of reverse magnetic fields},
  author={Taylor, J Brian},
  journal={Physical Review Letters},
  volume={33},
  number={19},
  pages={1139},
  year={1974},
  publisher={APS}
}

@ARTICLE{cozzo2023coronal,
       author = {{Cozzo}, G. and {Reid}, J. and {Pagano}, P. and {Reale}, F. and {Hood}, A.~W.},
        title = "{Coronal energy release by MHD avalanches. Effects on a structured, active region, multi-threaded coronal loop}",
      journal = {\aap},
         year = 2023,
        month = oct,
       volume = {678},
          eid = {A40},
        pages = {A40},
          doi = {10.1051/0004-6361/202346689},
archivePrefix = {arXiv},
       eprint = {2306.06047},
 primaryClass = {astro-ph.SR},
       adsurl = {https://ui.adsabs.harvard.edu/abs/2023A&A...678A..40C}
}

@article{cozzo2023asymmetric,
  title={Asymmetric Twisting of Coronal Loops},
  author={Cozzo, Gabriele and Pagano, Paolo and Petralia, Antonino and Reale, Fabio},
  journal={Symmetry},
  volume={15},
  number={3},
  pages={627},
  year={2023},
  publisher={MDPI}
}

@article{cozzo2024coronal,
	author = {{Cozzo}, G and {Reid}, J. and {Pagano}, P. and {Reale}, F. and {Testa}, P. and {Hood}, A. W. and {Argiroffi}, C. and {Petralia}, A. and {Alaimo}, E. and {D’Anca}, F. and {Sciortino}, L. and {Todaro}, M. and {Lo Cicero}, U. and {Barbera}, M. and {de Pontieu}, B. and {Martinez-Sykora}, J.},
	title = {Coronal energy release by MHD avalanches - II. EUV line emission from a multi-threaded coronal loop},
	DOI= "10.1051/0004-6361/202450644",
	url= "https://doi.org/10.1051/0004-6361/202450644",
	journal = {Astronomy \& Astrophysics},
	year = 2024,
	volume = 689,
	pages = "A184",
}

@article{reale20163d,
  title={3D MHD modeling of twisted coronal loops},
  author={Reale, Fabio and Orlando, Salvatore and Guarrasi, M and Mignone, A and Peres, Giovanni and Hood, AW and Priest, ER},
  journal={The Astrophysical Journal},
  volume={830},
  number={1},
  pages={21},
  year={2016},
  publisher={IOP Publishing}
}

@article{mignone2007pluto,
  title={PLUTO: a numerical code for computational astrophysics},
  author={Mignone, Andrea and Bodo, G and Massaglia, S and Matsakos, Titos and Tesileanu, O ea and Zanni, C and Ferrari, Anthony},
  journal={The Astrophysical Journal Supplement Series},
  volume={170},
  number={1},
  pages={228},
  year={2007},
  publisher={IOP Publishing}
}

@article{roe1986characteristic,
  title={Characteristic-based schemes for the Euler equations},
  author={Roe, Philip L},
  journal={Annual review of fluid mechanics},
  volume={18},
  number={1},
  pages={337--365},
  year={1986},
  publisher={Annual Reviews 4139 El Camino Way, PO Box 10139, Palo Alto, CA 94303-0139, USA}
}

@inproceedings{sweby1985high,
  title={High resolution TVD schemes using flux limiters},
  author={Sweby, Peter K},
  booktitle={Large-scale computations in fluid mechanics},
  year={1985}
}

@article{balsara1999staggered,
  title={A staggered mesh algorithm using high order Godunov fluxes to ensure solenoidal magnetic fields in magnetohydrodynamic simulations},
  author={Balsara, Dinshaw S and Spicer, Daniel S},
  journal={Journal of Computational Physics},
  volume={149},
  number={2},
  pages={270--292},
  year={1999},
  publisher={Elsevier}
}

@article{bradshaw2013influence,
  title={The influence of numerical resolution on coronal density in hydrodynamic models of impulsive heating},
  author={Bradshaw, Stephen J and Cargill, Peter J},
  journal={The Astrophysical Journal},
  volume={770},
  number={1},
  pages={12},
  year={2013},
  publisher={IOP Publishing}
}

@article{johnston2019fast,
  title={A fast and accurate method to capture the solar corona/transition region enthalpy exchange},
  author={Johnston, CD and Bradshaw, SJ},
  journal={The Astrophysical Journal Letters},
  volume={873},
  number={2},
  pages={L22},
  year={2019},
  publisher={IOP Publishing}
}

@article{johnston2020modelling,
  title={Modelling the solar transition region using an adaptive conduction method},
  author={Johnston, Craig D and Cargill, PJ and Hood, AW and De Moortel, Ineke and Bradshaw, SJ and Vaseekar, AC},
  journal={Astronomy \& Astrophysics},
  volume={635},
  pages={A168},
  year={2020},
  publisher={EDP Sciences}
}

@article{johnston2021fast,
  title={A fast multi-dimensional magnetohydrodynamic formulation of the transition region adaptive conduction (TRAC) method},
  author={Johnston, Craig D and Hood, Alan W and De Moortel, Ineke and Pagano, Paolo and Howson, Thomas A},
  journal={Astronomy \& Astrophysics},
  volume={654},
  pages={A2},
  year={2021},
  publisher={EDP Sciences}
}

@article{spitzer1953transport,
  title={Transport phenomena in a completely ionized gas},
  author={Spitzer Jr, Lyman and H{\"a}rm, Richard},
  journal={Physical Review},
  volume={89},
  number={5},
  pages={977},
  year={1953},
  publisher={APS}
}

@article{reale2007diagnostics,
  title={Diagnostics of stellar flares from X-ray observations: from the decay to the rise phase},
  author={Reale, Fabio},
  journal={Astronomy \& Astrophysics},
  volume={471},
  number={1},
  pages={271--279},
  year={2007},
  publisher={EDP Sciences}
}

@article{testa2020coronal,
  title={On the coronal temperature in solar microflares},
  author={Testa, Paola and Reale, Fabio},
  journal={The Astrophysical Journal},
  volume={902},
  number={1},
  pages={31},
  year={2020},
  publisher={IOP Publishing}
}

@article{lemen2012atmospheric,
  title={The atmospheric imaging assembly (AIA) on the solar dynamics observatory (SDO)},
  author={Lemen, James R and Title, Alan M and Akin, David J and Boerner, Paul F and Chou, Catherine and Drake, Jerry F and Duncan, Dexter W and Edwards, Christopher G and Friedlaender, Frank M and Heyman, Gary F and others},
  journal={Solar Physics},
  volume={275},
  pages={17--40},
  year={2012},
  publisher={Springer}
}

@article{boerner2012initial,
  title={Initial calibration of the atmospheric imaging assembly (AIA) on the solar dynamics observatory (SDO)},
  author={Boerner, Paul and Edwards, Christopher and Lemen, James and Rausch, Adam and Schrijver, Carolus and Shine, Richard and Shing, Lawrence and Stern, Robert and Tarbell, Theodore and Title, Alan and others},
  journal={The Solar Dynamics Observatory},
  pages={41--66},
  year={2012},
  publisher={Springer}
}

@ARTICLE{2025A&A...695A..40C,
       author = {{Cozzo}, G. and {Pagano}, P. and {Reale}, F. and {Testa}, P. and {Petralia}, A. and {Martinez-Sykora}, J. and {Hansteen}, V. and {De Pontieu}, B.},
        title = "{Coronal energy release by MHD avalanches: III. Identification of a reconnection outflow from a nanoflare}",
      journal = {\aap},
         year = 2025,
        month = mar,
       volume = {695},
          eid = {A40},
        pages = {A40},
          doi = {10.1051/0004-6361/202452426},
archivePrefix = {arXiv},
       eprint = {2502.01796},
 primaryClass = {astro-ph.SR},
       adsurl = {https://ui.adsabs.harvard.edu/abs/2025A&A...695A..40C}
}

@article{vaiana1973identification,
  title={Identification and analysis of structures in the corona from X-ray photography},
  author={Vaiana, GS and Krieger, AS and Timothy, AF},
  journal={Solar Physics},
  volume={32},
  pages={81--116},
  year={1973},
  publisher={Springer}
}

@article{reale2014coronal,
  title={Coronal loops: observations and modeling of confined plasma},
  author={Reale, Fabio},
  journal={Living Reviews in Solar Physics},
  volume={11},
  number={1},
  pages={1--94},
  year={2014},
  publisher={Springer}
}

@article{parker1988nanoflares,
  title={Nanoflares and the solar X-ray corona},
  author={Parker, Eugene N},
  journal={The Astrophysical Journal},
  volume={330},
  pages={474--479},
  year={1988}
}

@article{peter2014discovery,
  title={Discovery of the Sun's million-degree hot corona},
  author={Peter, Hardi and Dwivedi, Bhola N},
  journal={Frontiers in Astronomy and Space Sciences},
  volume={1},
  pages={2},
  year={2014},
  publisher={Frontiers Media SA}
}

@article{edlen1943deutung,
  title={Die Deutung der Emissionslinien im Spektrum der Sonnenkorona. Mit 6 Abbildungen.},
  author={Edl{\'e}n, Bengt},
  journal={Zeitschrift f{\"u}r Astrophysik, Vol. 22, p. 30},
  volume={22},
  pages={30},
  year={1943}
}

@article{grotrian1939sonne,
  title={Sonne und Ionosph{\"a}re},
  author={Grotrian, W},
  journal={Naturwissenschaften},
  volume={27},
  number={33},
  pages={555--563},
  year={1939},
  publisher={Springer}
}

@ARTICLE{Alfven1947,
       author = {{Alfv{\'e}n}, H.},
        title = "{Magneto hydrodynamic waves, and the heating of the solar corona}",
      journal = {\mnras},
         year = 1947,
        month = jan,
       volume = {107},
        pages = {211},
          doi = {10.1093/mnras/107.2.211},
       adsurl = {https://ui.adsabs.harvard.edu/abs/1947MNRAS.107..211A}
}

@article{klimchuk2015key,
  title={Key aspects of coronal heating},
  author={Klimchuk, James A},
  journal={Philosophical Transactions of the Royal Society A: Mathematical, Physical and Engineering Sciences},
  volume={373},
  number={2042},
  pages={20140256},
  year={2015},
  publisher={The Royal Society Publishing}
}

@article{parnell2012contemporary,
  title={A contemporary view of coronal heating},
  author={Parnell, Clare E and De Moortel, Ineke},
  journal={Philosophical Transactions of the Royal Society A: Mathematical, Physical and Engineering Sciences},
  volume={370},
  number={1970},
  pages={3217--3240},
  year={2012},
  publisher={The Royal Society Publishing}
}

@article{beveridge2003model,
  title={A model for elemental coronal flux loops},
  author={Beveridge, C and Longcope, DW and Priest, Eric Ronald},
  journal={Solar Physics},
  volume={216},
  pages={27--40},
  year={2003},
  publisher={Springer}
}

@article{klimchuk2008highly,
  title={Highly efficient modeling of dynamic coronal loops},
  author={Klimchuk, JA and Patsourakos, S and Cargill, PJ},
  journal={The Astrophysical Journal},
  volume={682},
  number={2},
  pages={1351},
  year={2008},
  publisher={IOP Publishing}
}

@article{vekstein2009probing,
  title={Probing nanoflares with observed fluctuations of the coronal EUV emission},
  author={Vekstein, G},
  journal={Astronomy \& Astrophysics},
  volume={499},
  number={1},
  pages={L5--L8},
  year={2009},
  publisher={EDP Sciences}
}

@article{testa2013observing,
  title={Observing coronal nanoflares in active region moss},
  author={Testa, Paola and De Pontieu, Bart and Mart{\'\i}nez-Sykora, Juan and DeLuca, Ed and Hansteen, Viggo and Cirtain, Jonathan and Winebarger, Amy and Golub, Leon and Kobayashi, Ken and Korreck, Kelly and others},
  journal={The Astrophysical Journal Letters},
  volume={770},
  number={1},
  pages={L1},
  year={2013},
  publisher={IOP Publishing}
}

@ARTICLE{Testa2020moss,
       author = {{Testa}, Paola and {Polito}, Vanessa and {De Pontieu}, Bart},
        title = "{IRIS Observations of Short-term Variability in Moss Associated with Transient Hot Coronal Loops}",
      journal = {\apj},
         year = 2020,
        month = feb,
       volume = {889},
       number = {2},
          eid = {124},
        pages = {124},
          doi = {10.3847/1538-4357/ab63cf},
archivePrefix = {arXiv},
       eprint = {1910.08201},
 primaryClass = {astro-ph.SR},
       adsurl = {https://ui.adsabs.harvard.edu/abs/2020ApJ...889..124T}
}

@ARTICLE{Cho2023,
       author = {{Cho}, Kyuhyoun and {Testa}, Paola and {De Pontieu}, Bart and {Polito}, Vanessa},
        title = "{A Statistical Study of IRIS Observational Signatures of Nanoflares and Nonthermal Particles}",
      journal = {\apj},
         year = 2023,
        month = mar,
       volume = {945},
       number = {2},
          eid = {143},
        pages = {143},
          doi = {10.3847/1538-4357/acb7da},
archivePrefix = {arXiv},
       eprint = {2211.06832},
 primaryClass = {astro-ph.SR},
       adsurl = {https://ui.adsabs.harvard.edu/abs/2023ApJ...945..143C}
}

@article{testa2014evidence,
  title={Evidence of nonthermal particles in coronal loops heated impulsively by nanoflares},
  author={Testa, Paola and De Pontieu, Bart and Allred, Joel and Carlsson, Mats and Reale, Fabio and Daw, Adrian and Hansteen, Viggo and Martinez-Sykora, Juan and Liu, Wei and DeLuca, EE and others},
  journal={Science},
  volume={346},
  number={6207},
  pages={1255724},
  year={2014},
  publisher={American Association for the Advancement of Science}
}

@article{gudiksen2005ab,
  title={An ab initio approach to the solar coronal heating problem},
  author={Gudiksen, Boris Vilhelm and Nordlund, {\AA}ke},
  journal={The Astrophysical Journal},
  volume={618},
  number={2},
  pages={1020},
  year={2005},
  publisher={IOP Publishing}
}

@article{golub2008x,
  title={The X-ray telescope (XRT) for the Hinode mission},
  author={Golub, L and Deluca, E and Austin, G and Bookbinder, J and Caldwell, D and Cheimets, P and Cirtain, J and Cosmo, M and Reid, P and Sette, A and others},
  journal={The Hinode Mission},
  pages={27--50},
  year={2008},
  publisher={Springer}
}

@article{hood2009coronal,
  title={Coronal heating by magnetic reconnection in loops with zero net current},
  author={Hood, AW and Browning, PK and Van der Linden, RAM},
  journal={Astronomy \& Astrophysics},
  volume={506},
  number={2},
  pages={913--925},
  year={2009},
  publisher={EDP Sciences}
}

@article{baty1996electric,
  title={Electric current concentration and kink instability in line-tied coronal loops.},
  author={Baty, H and Heyvaerts, J},
  journal={Astronomy and Astrophysics, v. 308, p. 935-950},
  volume={308},
  pages={935--950},
  year={1996}
}

@article{antolin2021reconnection,
  title={Reconnection nanojets in the solar corona},
  author={Antolin, Patrick and Pagano, Paolo and Testa, Paola and Petralia, Antonino and Reale, Fabio},
  journal={Nature Astronomy},
  volume={5},
  number={1},
  pages={54--62},
  year={2021},
  publisher={Nature Publishing Group UK London}
}

@article{de2022probing,
  title={Probing the physics of the solar atmosphere with the multi-slit solar explorer (MUSE). I. coronal heating},
  author={De Pontieu, Bart and Testa, Paola and Mart{\'\i}nez-Sykora, Juan and Antolin, Patrick and Karampelas, Konstantinos and Hansteen, Viggo and Rempel, Matthias and Cheung, Mark CM and Reale, Fabio and Danilovic, Sanja and others},
  journal={The Astrophysical Journal},
  volume={926},
  number={1},
  pages={52},
  year={2022},
  publisher={IOP Publishing}
}

@article{landi2013prominence,
  title={Prominence plasma diagnostics through extreme-ultraviolet absorption},
  author={Landi, E and Reale, F},
  journal={The Astrophysical Journal},
  volume={772},
  number={1},
  pages={71},
  year={2013},
  publisher={IOP Publishing}
}

@article{widing1992element,
  title={Element abundances and plasma properties in a coronal polar plume},
  author={Widing, KG and Feldman, Ur},
  journal={The Astrophysical Journal},
  volume={392},
  pages={715--721},
  year={1992}
}

@article{anders1989abundances,
  title={Abundances of the elements: Meteoritic and solar},
  author={Anders, Edward and Grevesse, Nicolas},
  journal={Geochimica et Cosmochimica acta},
  volume={53},
  number={1},
  pages={197--214},
  year={1989},
  publisher={Elsevier}
}

@article{arber2001staggered,
  title={A staggered grid, Lagrangian--Eulerian remap code for 3-D MHD simulations},
  author={Arber, TD and Longbottom, AW and Gerrard, CL and Milne, AM},
  journal={Journal of Computational Physics},
  volume={171},
  number={1},
  pages={151--181},
  year={2001},
  publisher={Elsevier}
}

@article{hudson2010observations,
  title={Observations of solar and stellar eruptions, flares, and jets},
  author={Hudson, Hugh},
  journal={Heliophysics: space storms and radiation: causes and effects. Cambridge University Press, London. p},
  volume={123},
  year={2010}
}

@article{drake1971characteristics,
  title={Characteristics of soft solar X-ray bursts},
  author={Drake, Jerry F},
  journal={Solar Physics},
  volume={16},
  pages={152--185},
  year={1971},
  publisher={Springer}
}

@article{crosby1993frequency,
  title={Frequency distributions and correlations of solar X-ray flare parameters},
  author={Crosby, Norma B and Aschwanden, Markus J and Dennis, Brian R},
  journal={Solar Physics},
  volume={143},
  pages={275--299},
  year={1993},
  publisher={Springer}
}

@article{dennis1985solar,
  title={Solar hard X-ray bursts},
  author={Dennis, Brian R},
  journal={Solar physics},
  volume={100},
  pages={465--490},
  year={1985},
  publisher={Springer}
}

@article{howson2022effects,
  title={The effects of driving time scales on coronal heating in a stratified atmosphere},
  author={Howson, Thomas Alexander and De Moortel, Ineke},
  journal={Astronomy \& Astrophysics},
  volume={661},
  pages={A144},
  year={2022},
  publisher={EDP Sciences}
}

@article{rosner1978dynamics,
  title={Dynamics of the quiescent solar corona},
  author={Rosner, Ro and Tucker, W Ho and Vaiana, GS},
  journal={Astrophysical Journal, Part 1, vol. 220, Mar. 1, 1978, p. 643-645, 647, 649-653, 655-665. Research supported by the Smithsonian Institution;},
  volume={220},
  pages={643--645},
  year={1978}
}

@article{landini1975loop,
  title={A loop model of active coronal regions},
  author={Landini, M and Monsignori Fossi, BC},
  journal={Astronomy and Astrophysics, vol. 42, no. 2, Aug. 1975, p. 213-220.},
  volume={42},
  pages={213--220},
  year={1975}
}

@article{guarrasi2010coronal,
  title={Coronal fuzziness modeled with pulse-heated multi-stranded loop systems},
  author={Guarrasi, Massimiliano and Reale, Fabio and Peres, Giovanni},
  journal={The Astrophysical Journal},
  volume={719},
  number={1},
  pages={576},
  year={2010},
  publisher={IOP Publishing}
}

@article{cargill2004nanoflare,
  title={Nanoflare heating of the corona revisited},
  author={Cargill, Peter J and Klimchuk, James A},
  journal={The Astrophysical Journal},
  volume={605},
  number={2},
  pages={911},
  year={2004},
  publisher={IOP Publishing}
}

@article{hood1979equilibrium,
  title={The equilibrium of solar coronal magnetic loops},
  author={Hood, AW and Priest, ER},
  journal={Astronomy and Astrophysics},
  volume={77},
  pages={233--251},
  year={1979}
}

@incollection{testa2023solar,
  title={The Solar X-ray Corona},
  author={Testa, Paola and Reale, Fabio},
  booktitle={Handbook of X-ray and Gamma-ray Astrophysics},
  pages={1--38},
  year={2023},
  publisher={Springer}
}

@article{de2019multi,
  title={The multi-slit approach to coronal spectroscopy with the multi-slit solar explorer (MUSE)},
  author={De Pontieu, Bart and Mart{\'\i}nez-Sykora, Juan and Testa, Paola and Winebarger, Amy R and Daw, Adrian and Hansteen, Viggo and Cheung, Mark CM and Antolin, Patrick},
  journal={The Astrophysical Journal},
  volume={888},
  number={1},
  pages={3},
  year={2019},
  publisher={IOP Publishing}
}

@article{schindler1988general,
  title={General magnetic reconnection, parallel electric fields, and helicity},
  author={Schindler, K and Hesse, M and Birn, J},
  journal={Journal of Geophysical Research: Space Physics},
  volume={93},
  number={A6},
  pages={5547--5557},
  year={1988},
  publisher={Wiley Online Library}
}

@article{hesse1988theoretical,
  title={A theoretical foundation of general magnetic reconnection},
  author={Hesse, M and Schindler, K},
  journal={Journal of Geophysical Research: Space Physics},
  volume={93},
  number={A6},
  pages={5559--5567},
  year={1988},
  publisher={Wiley Online Library}
}

@article{johnston2025self,
  title={Self-Consistent Heating of the Magnetically Closed Solar Corona: Generation of Nanoflares, Thermodynamic Response of the Plasma and Observational Signatures},
  author={Johnston, Craig D and Daldorff, Lars KS and Klimchuk, James A and Mondal, Shanwlee Sow and Barnes, Will T and Leake, James E and Reid, Jack and Parker, Jacob D},
  journal={arXiv preprint arXiv:2508.12952},
  year={2025}
}

@article{del2021chianti,
  title={CHIANTI—an atomic database for emission lines. XVI. Version 10, further extensions},
  author={Del Zanna, G and Dere, KP and Young, PR and Landi, E},
  journal={The Astrophysical Journal},
  volume={909},
  number={1},
  pages={38},
  year={2021},
  publisher={IOP Publishing}
}

@article{feldman1992elemental,
  title={Elemental abundances in the upper solar atmosphere},
  author={Feldman, Uri},
  journal={Physica Scripta},
  volume={46},
  number={3},
  pages={202},
  year={1992},
  publisher={IOP Publishing}
}

@article{poduval2013point,
  title={Point-spread functions for the extreme-ultraviolet channels of SDO/AIA telescopes},
  author={Poduval, B and DeForest, CE and Schmelz, JT and Pathak, S},
  journal={The Astrophysical Journal},
  volume={765},
  number={2},
  pages={144},
  year={2013},
  publisher={IOP Publishing}
}

@article{sen2025merging,
  title={Merging plasmoids and nanojet-like ejections in a coronal current sheet},
  author={Sen, Samrat and Moreno-Insertis, Fernando},
  journal={Astronomy \& Astrophysics},
  volume={699},
  pages={A106},
  year={2025},
  publisher={EDP Sciences}
}

@article{guarrasi2014mhd,
  title={MHD modeling of coronal loops: the transition region throat},
  author={Guarrasi, M and Reale, F and Orlando, S and Mignone, A and Klimchuk, JA},
  journal={Astronomy \& Astrophysics},
  volume={564},
  pages={A48},
  year={2014},
  publisher={EDP Sciences}
}

@article{Bonet2008,
  author       = {Bonet, J.\,A. and Márquez, I. and Sánchez Almeida, J. and Cabello, I. and Domingo, V.},
  title        = {Convectively Driven Vortex Flows in the Sun},
  journal      = {The Astrophysical Journal Letters},
  volume       = {687},
  pages        = {L131},
  year         = {2008},
  doi          = {10.1086/593329}
}

@article{Galsgaard1996,
  author       = {Galsgaard, K. and Nordlund, Å.},
  title        = {Heating and Activity of the Solar Corona — 1. Boundary Shearing of an Initially Homogeneous Magnetic Field},
  journal      = {Journal of Geophysical Research},
  volume       = {101},
  pages        = {13\,445},
  year         = {1996},
  doi          = {10.1029/96JA00428}
}

@article{Hermans2021A&A655A36,
  author       = {Hermans, J. and Keppens, R.},
  title        = {Cooling and heating timescales in magnetised coronal loops},
  journal      = {Astronomy \& Astrophysics},
  volume       = {655},
  pages        = {A36},
  year         = {2021},
  doi          = {10.1051/0004-6361/202140958}
}

@article{Sen2024A&A688A64,
  author       = {Sen, S. and Keppens, R. and Hermans, J. and Zhou, Y.},
  title        = {Coronal rain from magnetohydrodynamic simulations: Timescales of cooling and condensations},
  journal      = {Astronomy \& Astrophysics},
  volume       = {688},
  pages        = {A64},
  year         = {2024},
  doi          = {10.1051/0004-6361/202348031}
}

@article{Sen2025A&A699A106,
  author       = {Sen, S. and Moreno-Insertis, F.},
  title        = {Thermal evolution and characteristic cooling timescales of reconnecting coronal loops},
  journal      = {Astronomy \& Astrophysics},
  volume       = {699},
  pages        = {A106},
  year         = {2025},
  doi          = {10.1051/0004-6361/202449010}
}
\bibliographystyle{aasjournalv7}

\end{document}